\documentclass[twocolumn,amsmath,amssymb,prl]{revtex4}
\usepackage{latexsym}
\usepackage{graphicx}
\usepackage{color}
\usepackage{amsmath}

\usepackage{xcolor}
\usepackage{textcomp,gensymb}
\usepackage{longtable,threeparttablex,booktabs}
\usepackage{array}

\renewcommand{\thefigure}{\textbf{\arabic{figure}}}
\setcitestyle{super,open={},close={}}

\makeatletter
\renewcommand\@biblabel[1]{#1.}
\makeatother

\begin{document}
\title{Observation of Distinct Superconducting Phases in Hyperdoped p-type Germanium}

\author{Kasra Sardashti$^{1}$}
\author{Tri D. Nguyen$^{1,2}$}
\author{Wendy L. Sarney$^{3}$}
\author{Asher C. Leff$^{3}$}
\author{Mehdi Hatefipour$^{1}$}
\author{Matthieu~C.~Dartiailh$^{1}$}
\author{Joseph Yuan$^{1}$}
\author{William Mayer$^{1}$}
\author{Javad Shabani$^{1*}$}

\affiliation{$^{1}$Center for Quantum Phenomena, Department of Physics, New York University, NY 10003, USA\\
$^{2}$Department of Physics, City College of New York, City University of New York, NY 10031, USA\\
$^{3}$ CCDC U.S. Army Research Laboratory, Adelphi, MD 20783 USA
}

\date{\today}

\maketitle

{\bf Realization of superconductivity in Group IV semiconductors could have a strong impact in the direction quantum technologies will take in the future. Therefore, it is imperative to understand the nature of the superconducting phases in materials such as Silicon and Germanium. Here, we report systematic synthesis and characterization of superconducting phases in hyperdoped Germanium prepared by Gallium ion implantation beyond its solubility limits. The resulting structural and physical characteristics have been tailored by changing the implantation energy and activation annealing temperature. Surprisingly, in addition to the poly-crystalline phase with weakly-coupled superconducting Ga clusters we find a nano-crystalline phase with quasi-2D characteristics consisting of a thin Ga film constrained near top surfaces. The new phase shows signatures of strong disorder such as anomalous B${\rm c}$ temperature dependence and crossings in magentoresistance isotherms. Apart from using hyperdoped Ge as a potential test-bed for studying signatures of Quantum phase transitions (e.g. quantum Griffith singularity), our results suggest the possibility of integration of hyperdoped Ge nano-crystalline phase into superconducting circuits due to its 2D nature. }\\

\noindent
\textbf{Introduction}

The last decade has witnessed significant advances in the materials synthesis for superconducting circuits. Amongst them, superconductors--semiconductor (S-Sm) hybrid materials platforms have attracted a great deal of attention. For instance, many exciting avenues, built upon the success of gate-tunable hybrid epitaxial Al--InAs Josephson junctions technology \cite{shabani_two-dimensional_2016, krogstrup_epitaxy_2015, mayer_superconducting_2019}, have been explored including gate-tunable transmon qubits \cite{casparis_superconducting_2018}, gate-tunable quantum busses \cite{casparis_voltage-controlled_2019}, and platforms for superconducting spintronics and topological superconducting qubits \cite{mayer_gate_2020, fornieri_evidence_2019,  mayer2019phase}. However, Group IV semiconductors are of particular interest for integration into Sm-S platforms due to their high purity and compatibility with highly scalable complimentary metal-oxide-semiconductor (CMOS) technologies \cite{shim_bottom-up_2014, shim_superconducting-semiconductor_2015}. Realization of superconductivity by hyperdoping is believed to facilitate such integration. So far, superconductivity has been successfully documented for diamond \cite{ekimov_superconductivity_2004, okazaki_signature_2015}, Silicon \cite{bustarret_superconductivity_2006, skrotzki_-chip_2010, thorgrimsson_effect_2020} and Germanium \cite{herrmannsdorfer_superconducting_2009, fiedler_superconducting_2012} by hyperdoping with acceptors (i.e. B and Ga). The high doping levels are expected to surpass the metal--insulator transition limits, leading to a narrower band gap and larger inter-valley coupling \cite{cohen_superconductivity_1964, blase_superconducting_2009}.

Germanium (Ge) is a compelling candidate for S-Sm platforms because ultra-clean materials with high hole mobilities may be achieved \cite{scappucci_germanium_2020}. In its pure and alloyed state, Ge is used in a wide variety of room-temperature electronics with high integrability with Si technology. Superconductivity in Ge is demonstrated through Ga$^+$ ion implantation followed by dopant activation via rapid thermal annealing (RTA) or flash lamp annealing (FLA) at temperatures near the Ge melting point ($\sim 938 ^{\circ}C$) \cite{fiedler_superconducting_2011, herrmannsdorfer_superconducting_2009, skrotzki_impact_2011, prucnal_superconductivity_2019}. Such high temperatures were intended for recovering the crystallinity of the Ge substrate damaged by implantation. Nevertheless, given its low solubility in Ge (maximum of $\sim$ 1.1\% at about 700 $^{\circ}$C \cite{olesinski_gage_1985}), Ga is expected to not only take interstitial sites, but to segregate into clusters within the bulk or accumulate near the top surfaces. While such phase may not be thermodynamically stable, rapid cooling after annealing could help them reside within the implanted regions as metastable phases.

\setcitestyle{numbers}
\begin{figure*}[ht!]
  \includegraphics[width=0.8\textwidth]{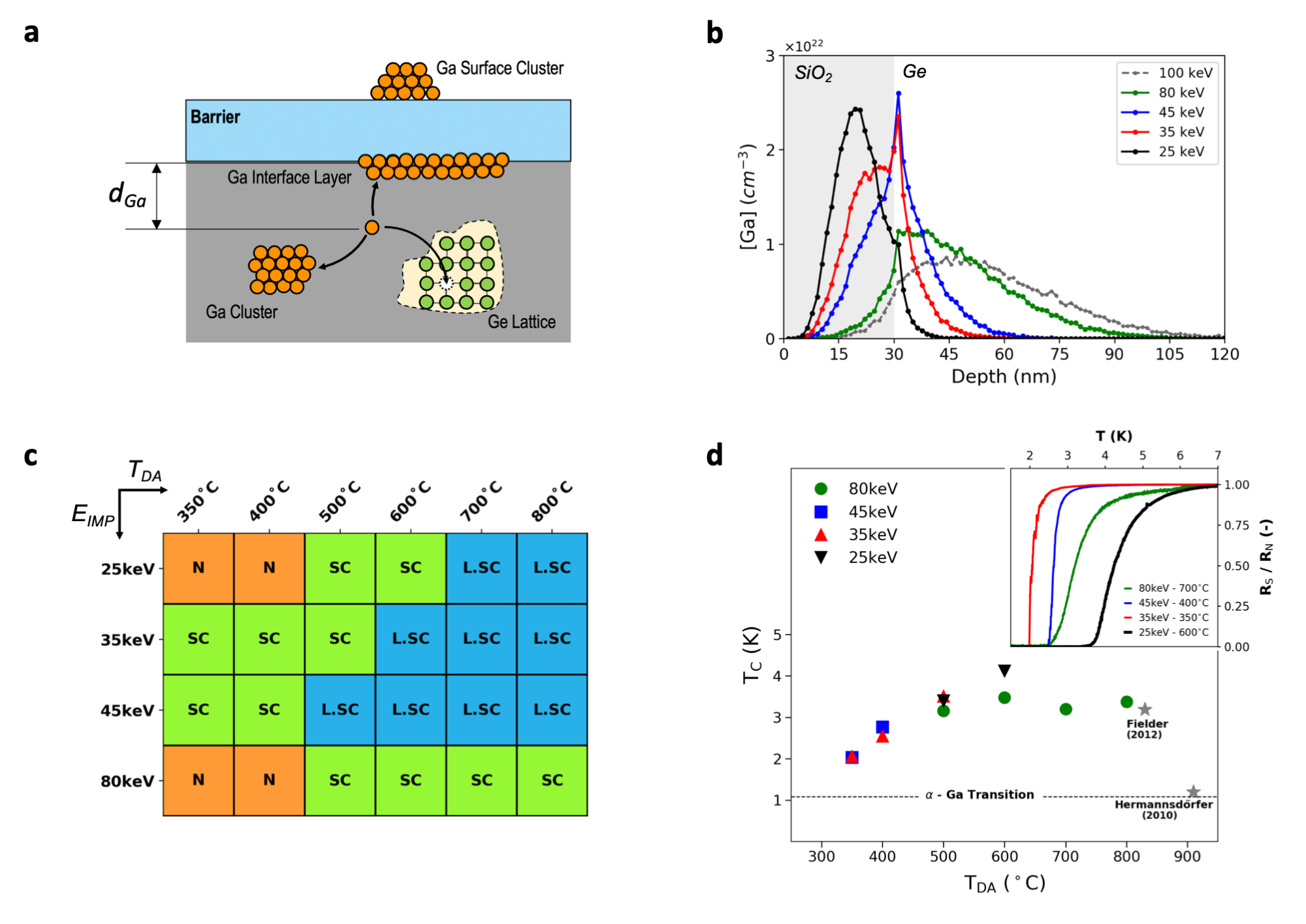}
    \caption{\label{Fig.1} \textbf{a} A schematic detailing the various pathways implanted Ga atoms can take within the Ge matrix, including substituting Ge as a dopant, cluster formation within the implanted region, accumulation at the $SiO_2/Ge$ interface, and finally diffusion through the $SiO_2$ to form Ga droplets on the surface. \textbf{b} One-dimensional simulation of Ga concentration vs. depth for various implantation energies from 25~keV to 100~keV, calculated by TRIM for $Ga^+$ fluence of $4 \times 10^{16}\; cm^{-2}$. \textbf{c} Summary of transport properties (marked by "N"=normal, "SC"=superconducting, and "L.SC"=locally superconducting) vs. processing conditions used in this study i.e. implantation energies and activation annealing temperatures. \textbf{d} Superconducting transition temperatures vs. anneal temperature for "SC" samples. The reference T$_{c}$ data points (denoted by $\star$) were adapted from Ref.\cite{herrmannsdorfer_superconductivity_2010, fiedler_superconducting_2012}. The $\alpha$-Ga  T$_{c}$ value is adapted from Ref.\cite{gregory_superconducting_1966}. The inset shows the transition to zero-resistance state as a function of temperature for four samples: one from each implantation energy.}
\end{figure*}
\setcitestyle{super}

In this work, we show the possibility to tune the superconducting properties of hyperdoped Ge using two major processing parameters: 1) Ga$^+$ implantation energy (E$_{\rm IMP}$); 2) dopant activation anneal temperature (T$_{\rm DA}$). We show superconductivity in hyperdoped Ge over a wide parameter space with E$_{\rm IMP}$ and T$_{\rm DA}$ as low as 25~keV and 350 $^{\circ}$C, respectively. In this extended parameter space, two structurally and physically distinct phases are identified, namely the poly-crystalline (PC) and the nano-crystalline (NC) phases. The PC phase has low critical magnetic fields (B$_{\rm c}$) with a 3D network of coupled superconducting Ga puddles \cite{spivak_theory_2008}. Conversely, the NC phase superconductivity, stemming from Ga thin film accumulated near the SiO$_{\rm 2}$/Ge interface, can survive in-plane fields as high as $\sim$ 8~T. Interestingly, samples from both PC and NC phases are found to have anomalous temperature dependence of B$_{\rm c}$ as T$\rightarrow$0, pointing to the presence of quenched disorder and vortex glass state \cite{sacepe_low-temperature_2019}. Moreover, clear crossing zones are identified for sheet resistance, R$_{\rm s}$(B), curves in 2D NC phases, offering the possibility of using this system to study properties of quantum critical point (QCP), and in general quantum phase transitions (QPTs) \cite{xing_quantum_2015, saito_quantum_2018}. Signatures of high disorder could be applicable to quantum circuits where high kinetic inductance is needed. This could invoke a range of applications as in superinductors \cite{hazard_nanowire_2019}, magnetic-field-resistant superconducting resonators \cite{niepce_high_2019, samkharadze_high-kinetic-inductance_2016} and phase slip qubits \cite{peltonen_hybrid_2018, astafiev_coherent_2012}.\\

\noindent
\textbf{Results}

\textbf{Transport properties vs. processing conditions.} Figure \ref{Fig.1}a displays the pathways Ga ions may take within the Ge substrate during the activation annealing. In this picture, an oxide barrier is depicted on top of the Ge substrate, as it is commonly used to prevent surface damage during the ion implantation process. Due to low solubility, beside being incorporated into the bulk Ge lattice as a dopant, Ga could segregate into clusters within the implanted region, form a thin layer at the Ge/barrier interface or even diffuse through the barrier to form surface droplets. The percentage of Ga atoms participating in each process depends on the average distance of the implanted atoms from the top surface as well as the amount thermal energy provided during the annealing. Figure \ref{Fig.1}b shows the Ga distribution as a function of E$_{\rm IMP}$, simulated using the Transport of Ions In Matter (TRIM) Monte Carlo software \cite{ziegler_stopping_2013}. The lines show Ga concentration (per cm$^{\rm 3}$) vs. depth for E$_{\rm IMP}$ = 25, 35, 45, and 80~keV all experimentally implemented in this study. The grey line represents Ga distribution for E$_{\rm IMP}$ = 100~keV, serving as a reference to the previous reports on superconducting Ge \cite{fiedler_superconducting_2011, herrmannsdorfer_superconducting_2009, heera_depth-resolved_2014}.

\begin{figure*}[ht!]
  \includegraphics[width=0.8\textwidth]{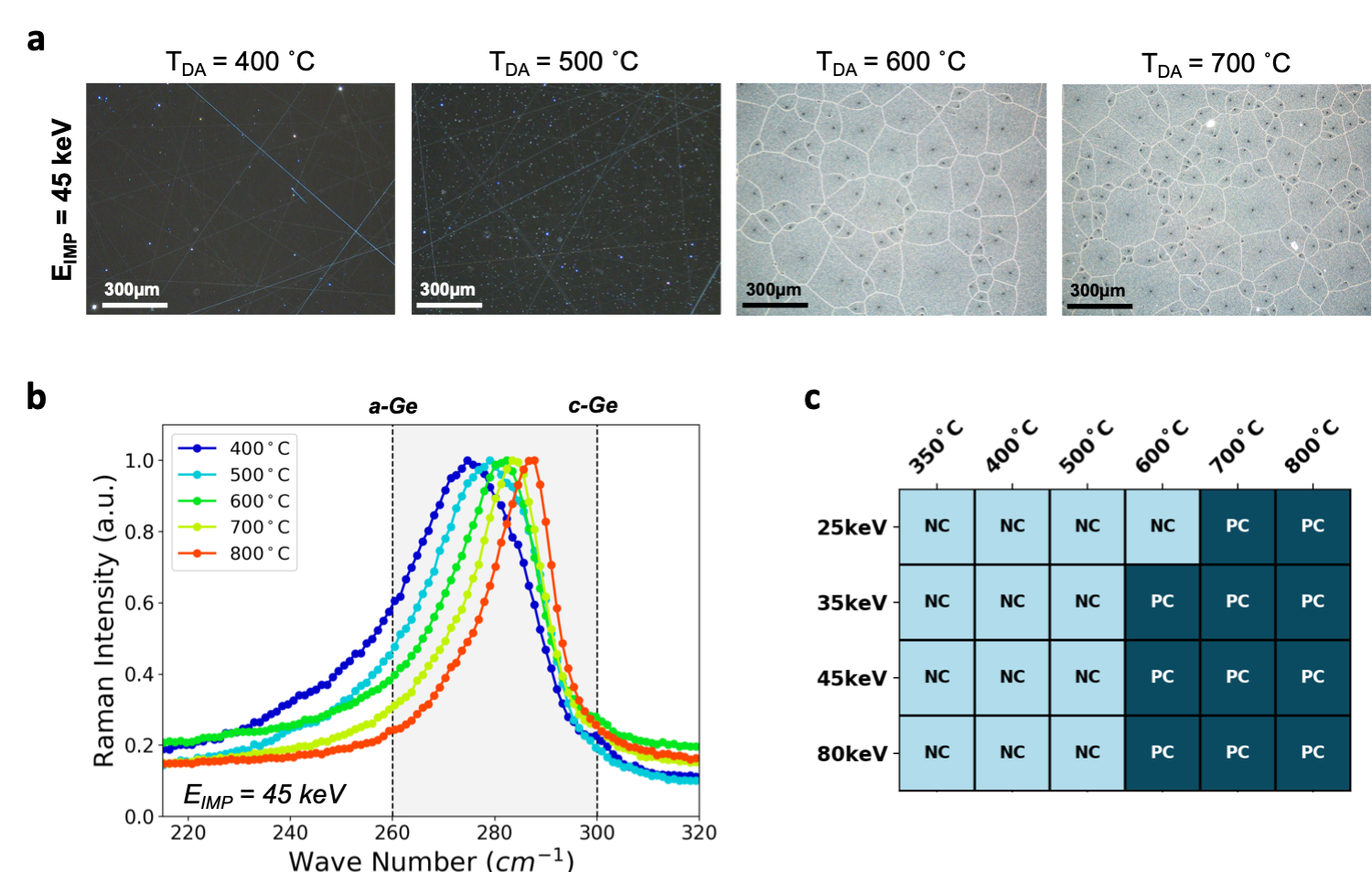}
      \caption{\label{Fig.2} Structural evolution of the Ga-implanted Ge films with $_{\rm IMP}$ = 45~keV. \textbf{a} Dark field optical images of four samples annealed at 400~$^{\circ}$C--700~$^{\circ}$C. \textbf{b} Raman peaks shift for samples annealed at 400~$^{\circ}$C--800~$^{\circ}$C. The standard peak positions for amorphous (a-Ge) and single crytsalline Ge (c-Ge) are marked by vertical dashed lines.  \textbf{c} An overview of sample phase vs. E$_{\rm IMP}$ and T$_{\rm DA}$, for all 24 processing conditions tested in this study. "PC" \& "NC" denote poly-crystalline and nano-crystalline, respectively.  }
\end{figure*}

Implanted samples underwent RTA for 1~min in order to partially recover the crystal structure and to incorporate Ga atoms into interstitial sites as acceptors. RTA was carried out in a wide temperature range from 300~$^{\circ}$C to 800~$^{\circ}$C. Figure \ref{Fig.1}c shows a matrix of E$_{\rm IMP}$ vs. T$_{\rm DA}$ for the Ga-implanted Ge samples prepared in this study. The boxes labeled by "SC" signify samples with clear superconducting transitions to a zero-resistance phase (see inset of \ref{Fig.1}d). Samples labeled with "N" showed non-zero resistance all the way down to 1.5~K in the absence of any distinctive superconducting transition above that temperature. Finally, samples labeled by "L.SC" that stands for localized superconductivity displayed a clear drop in resistance at 5--7~K, yet finite resistance at 1.5 K (see supplementary figure \ref{Fig.S1}). Those systems are believed to have superconducting puddles that are too widely separated by normal phases,relative to the correlation lengths necessary for their coupling \cite{deutscher_upper_1982, john_phase_1986}. It should be noted that observing a zero-resistance state at all implantation energies explored in this study, and temperatures as low as 350 $^{\circ}$C, is a significant expansion of the processing parameter space proposed in previous reports (i.e. 100keV, 700-910$^{\circ}$C) \cite{fiedler_superconducting_2011, fiedler_superconducting_2012, herrmannsdorfer_superconducting_2009}.

Dependence of superconducting transition temperature ($T_c$) on E$_{\rm IMP}$ and T$_{\rm DA}$, for the samples with complete transition to zero-resistance phase (i.e. "SC"), is shown in Figure \ref{Fig.1}d. The T$_{\rm c}$ values were extracted from sheet resistance vs. temperature curves at points where resistance is 50\% of normal resistance at 10 K. $T_c$ is relatively constant within the margins of error for the sample with E$_{\rm IMP}$ = 80~keV. By reducing E$_{\rm IMP}$ to 45 and 35~keV, where Ga peak is still within the Ge substrate (see Figure \ref{Fig.1}b), $T_c$ is lowered to minimum value of 2~K while the normal-superconductor transition became sharper (see supplementary figure \ref{Fig.S2} for an overview transition widths). However, for E$_{\rm IMP}$ = 25~keV, the T$_c$ one again raises to 3.5 and 4.0~K for the two superconducting samples with T$_{\rm DA}$ = 500 $^{\circ}$C and 600 $^{\circ}$C. $T_c$ of 4.0 K, is $\sim$ 1 K above the record value reported in the literature for hyper Ga-doped Ge for R$_{\rm c}$ = 0.5 R$_{\rm n}$. Similarly, the transition widths for for samples with E$_{\rm IMP}$ = 25~keV approach the values for E$_{\rm IMP}$ = 80~keV. At such low E$_{\rm IMP}$s, as the ion beam spreads over a smaller interaction volume and the majority Ga stays inside the barrier, significant mixing between Ge, Ga and SiO$_2$ is expected.  Therefore, a higher density of structural defects may be expected in the 25~keV samples, leading to more non-trivial transitional behavior.

\textbf{Structural and compositional characterization.} Variations in the E$_{\rm IMP}$ and T$_{\rm DA}$ have significant impact on the microstructure of the films. Figure \ref{Fig.2}a shows the dark field (DF) optical images of the films prepared with  E$_{\rm IMP}$ = 45~keV and T$_{\rm DA}$ = 400--700~$^{\circ}$C. From these DF images it is clear that between 500~$^{\circ}$C and 600~$^{\circ}$C, a transition occurs from crossing lines patterns to a poly-crystalline structure with a few hundred $\mu m$ wide grains. This behavior has been more or less observed at all implantation energies tested in this study (see figure \ref{Fig.S3}). Nonetheless, the average grain size for the surfaces with E$_{\rm IMP}$ = 25~keV is much smaller than other energies, leading to DF images with no micron-size features. In addition, the crossing lines are only present for E$_{\rm IMP} <$  80~keV. These lines were confirmed to be actual topographical features a few $nm$ depth as confirmed by atomic force microscopy measurements (see supplementary figures \ref{Fig.S4} and \ref{Fig.S5}). 

The crystallinity of the samples were further investigated by monitoring the Ge Raman peak position as a function of T$_{\rm DA}$, as illustrated in figure \ref{Fig.2}b. While implantation process led to near-complete amorphization of the top surface (spectrum not shown here), increasing the annealing temperature coincides with the Ge Raman peak shift from near amorphous Ge (a-Ge) to values closer to crystalline Ge (c-Ge) \cite{johnson_solid_2014, olivares_raman_2000}. This trend of increasing crystallinity with T$_{\rm DA}$ was universally seen across all implantation energies (see supplementary figure \ref{Fig.S5}). However, for lower shallower Ga$^{\rm +}$ implantation (i.e. E$_{\rm IMP}$ = 25 and 35~keV), especially at lower annealing temperature, secondary peak is noticed at 300 cm$^{\rm -1}$, consistent with signal detected from the c-Ge below the implanted region.

\begin{figure}[ht!]
  \includegraphics[width=0.49\textwidth]{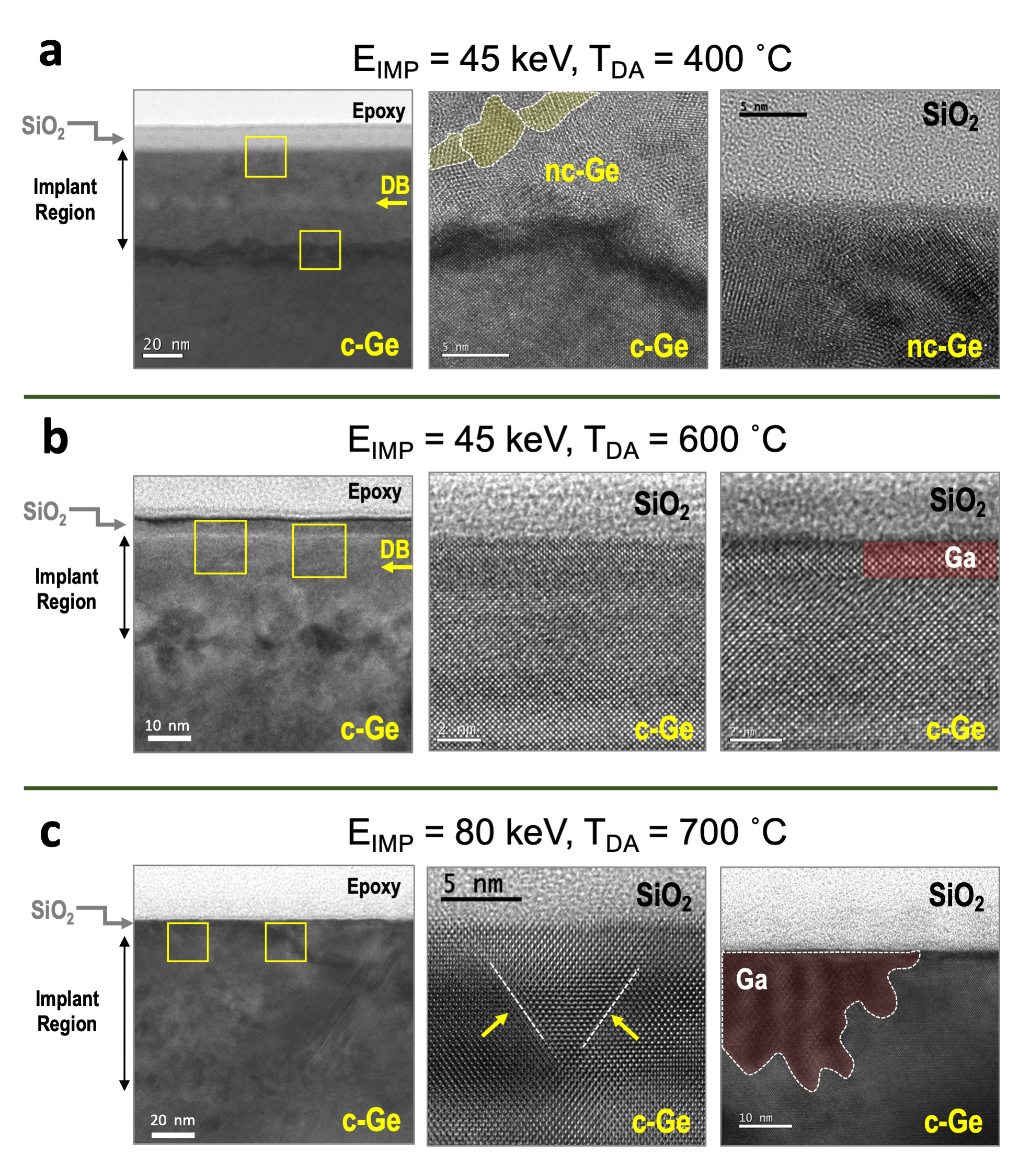}
      \caption{\label{Fig.3} TEM images of cross-sections prepared on three heavily Ga-doped Ge samples with: \textbf{a} E$_{\rm IMP}$ = 45 keV \& T$_{\rm DA}$ = 400 $\degree$C. \textbf{b} E$_{\rm IMP}$ = 45 keV \& T$_{\rm DA}$ = 600 $\degree$C. \textbf{c} E$_{\rm IMP}$ = 80 keV \& T$_{\rm DA}$ = 700 $\degree$C. The yellow boxes show in lower magnification images (left column) are visual guides to outline approximate areas from which the higher magnification images (middle and right columns) were taken. In (a) \& (b) disturbed bands, labeled by arrows and "DB" are seen a result of $SiO_2$ recoil during the implantation process. In (c) section, arrows denote stacking faults and twin boundary defects within the Ge after dopant activation annealing.}
\end{figure}

\noindent
Figure \ref{Fig.2}c summarizes our findings on the structure of the sample as a function of implantation energy and activation annealing temperature. Based on the DF optical images and Raman peak position, samples were divided into two general groups: i) Nano-crystalline (NC) with dark DF images and Raman shift closer to a-Ge; ii) Poly-crystalline (PC) with clear grain structure visible to optical microscopy and Raman peaks near c-Ge. Comparing this table with figure \ref{Fig.1}c, we note that for E$_{\rm IMP} <$  80~keV, only nano-crystalline films are superconducting with a zero-resistance state above 1.5~K. PC surfaces at those energies only "L.SC" with transitions around T = 6 K (near T$_{\rm c}$ for $\beta$-Ga \cite{campanini_raising_2018} and Ga:Si\cite{skrotzki_-chip_2010}). Instead, when E$_{\rm IMP}$ = 80~keV, the superconducting behavior spans from NC (T$_{\rm DA}$ = 500 $^{\circ}$C) to PC (T$_{\rm DA} \geq$ 600 $^{\circ}$C). Considering that the NC-PC transition occurs at relatively similar temperatures for all implantation energies, this difference in superconductivity may be attributed to the average depth distribution of Ga within the sample. For shallower Ga peaks, while higher anneal temperature successfully recovers parts of the surface crystallinity, it leads to a larger loss of Ga from the top surface into surface droplets or Ga vapor. For the samples with E$_{\rm IMP}$=80 keV, the average Ga atoms lie further from the top surface Ga clustering within the implanted region or near the SiO$_{\rm 2}$/Ge may be kinetically more favorable.

\begin{figure*}[ht!]
    \includegraphics[width=0.85\textwidth]{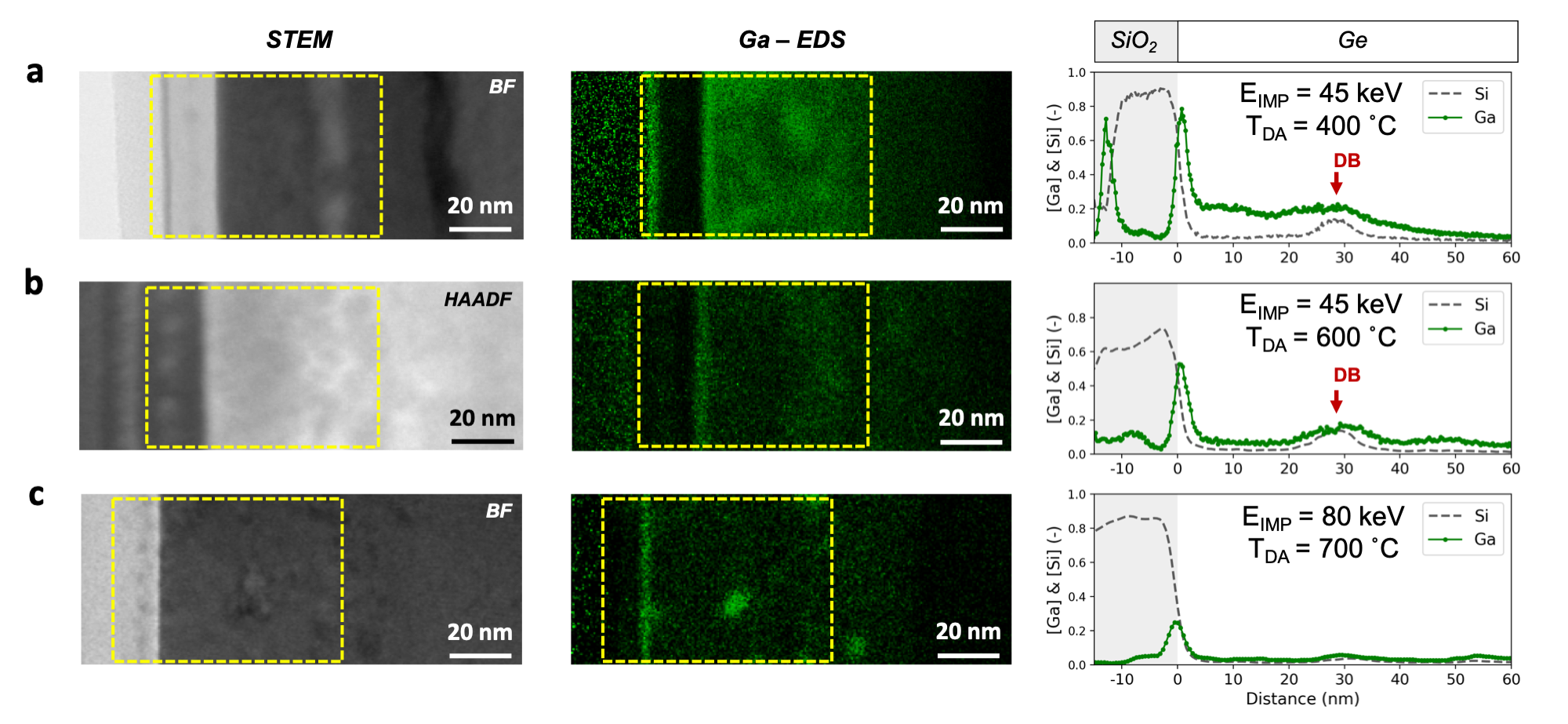}
      \caption{\label{Fig.4} STEM images and EDS elemental Ga maps for cross-sections of three Ga-implanted Ge samples. \textbf{a} E$_{\rm IMP}$ = 80 keV \& T$_{\rm DA}$ = 700 $\degree$C; \textbf{b} E$_{\rm IMP}$ = 45 keV \& T$_{\rm DA}$ = 400 $\degree$C; \textbf{c} E$_{\rm IMP}$ = 45 keV \& T$_{\rm DA}$ = 600 $\degree$C. The line traces for [Ga] (solid green lines) and [Si] (dotted grey lines), normalized to [Si]+[Ge], are shown on the right. The traces are averaged over the width of the areas outlined by the dotted rectangles in STEM images and maps. The line traces are aligned by setting the $SiO_2/Ge$ interface to be at zero distance. Disturbed bands are marked by red arrows and "DB".}
\end{figure*}

Structures of the implanted samples after dopant activation annealing were examined more closely by Transmission Electron Microscopy (TEM). As shown in Figure \ref{Fig.3}, a select number of samples with clear NC and PC structures were characterized by TEM. These are representative systems for shallow and deep Ga$^{\rm +}$ implantation. The samples with E$_{\rm IMP}$ = 45 keV (shown in figure \ref{Fig.3}a and \ref{Fig.3}b) have a disturbed band $\sim$ 25~nm below SiO$_{\rm 2}$ barrier due to $Si$ \& $O$ recoil during the implantation process. Focusing on the Ge surrounding the disturbed band, annealing at 400 $^{\circ}$C forms a multi-crystalline film with a few $nm$ wide grains (NC phase). In contrary, annealing at 600 $^{\circ}$C helped recover the crystallinity to a significant level, and allowed formation of a few monolayers thick crystalline Ga at the SiO$_{\rm 2}$/Ge interface (highlighted in red, right column in figure \ref{Fig.3}b). Such crystalline Ga layer appears to be discontinuous as evidenced by its absence in other high-resolution images of theSiO$_{\rm 2}$/Ge interface (middle column in figure \ref{Fig.3}b). This is consistent with the observation of a 6-K transition ($\beta$-Ga transition\cite{campanini_raising_2018}) to a non-zero resistance state. The difference in the crystallinity of these two systems was confirmed by electron diffraction measurements (see figure \ref{Fig.S7}).

Figure \ref{Fig.3}c displays the cross-sections of a Ge sample with E$_{\rm IMP}$ = 80 keV and T$_{\rm DA}$ = 700 $^{\circ}$C. Compared to the samples with E$_{\rm IMP}$ = 45 keV, the depth of the region disturbed by the implanting ions appears to be more than 2x larger. In addition, no visible sign of a disturbed band is present below the SiO$_{\rm 2}$ barrier. Under these conditions, the implanted region appears to consist of highly crystalline Ge in the 15-20~nm lenght scale, although with imperfections such as stacking faults (marked by dotted lines in figure \ref{Fig.3}c, middle column). Besides, the annealing temperature tended to be high enough to enable formation of large crystalline Ga puddles ($\geq$ 25 $nm$ wide) within the implanted region adjacent to the SiO$_{\rm 2}$/Ge interface (right-most column in figure \ref{Fig.3}c).

To better determine the Ga distribution in the samples, TEM measurements were complemented by Scanning Transmission Electron Microscopy (STEM) and Energy dispersive Spectroscopy (EDS) compositional mapping. Figure \ref{Fig.4} shows the STEM images, Ga elemental maps, and line traces of normalized Si \& Ga concentrations (i.e.  [Ga] =  Ga\%/ (Si\%+Ge\%)) for the three samples discussed above. For samples with E$_{\rm IMP}$ = 45 keV (figure \ref{Fig.4}a and b), once again disturbed bands were seen at $\sim$ 25-30 $nm$ below the SiO$_{\rm 2}$ interfaces, with excess amounts Si, O and Ga (see supplementary figure \ref{Fig.S8} for Ge and O line scans). For the sample with E$_{\rm IMP}$ = 80 keV shown in figure \ref{Fig.4}c, only small Ga-rich clusters were observed, confirming that disturbed band is only a product of low-energy implantation.

At E$_{\rm IMP}$ = 45~keV \& T$_{\rm DA}$ = 400~$^{\circ}$C, relatively large [Ga] ($\geq$ 20\%) are trapped within the implanted region, particularly in between the SiO$_{\rm 2}$ and the disturbed band (figure \ref{Fig.4}a). Additionally,sharp Ga peaks were seen both at the SiO$_{\rm 2}$/Ge interface and SiO$_{\rm 2}$ top surfaces. With raising the annealing temperature to 600~$^{\circ}$C, the Ga atoms seem to diffuse through the structure both towards the Ge substrate and the top barrier (figure \ref{Fig.4}b). This is evidenced by reduced [Ga] within the implanted region (including interfaces and disturbed band), and enhanced [Ga] below the band. These results are also consistent with the high-resolution TEM images in figure \ref{Fig.3}b, where there is no abrupt boundary between implanted region and Ge substrate. Elemental mapping for the Ge sample with E$_{\rm IMP}$ = 80 keV \& T$_{\rm DA}$ = 700 $^{\circ}$C (shown in figure \ref{Fig.4}c) displays small [Ga] ($\leq$ 5\%) within the implanted region, only concentrated in clusters that are dispersed across the cross-section, with spacing in the order of 20~$nm$. Moreover, Ga accumulates at the SiO$_{\rm 2}$/Ge interfaces into a thin layer with a peak relative concentration of $\sim$ 25\% (more than 2x smaller peaks in samples with E$_{\rm IMP}$ = 45keV). For samples with E$_{\rm IMP}$ = 80keV, the amount of Ga within the bulk of the implanted region and relative height of the interface Ga peak may also be controlled by varying T${\rm DA}$ (see supplementary information figure \ref{Fig.S8}a \& b, figure \ref{Fig.S9}).

Another piece of information that may be crucial in explaining the role of Ga distribution in the superconductivity of the PC and NC phases is the amount of Ga incorporated into the Ge lattice as a dopant. Therefore, we measured the hole density (n$_{\rm h}$) for several samples with various E$_{\rm IMP}$s. Details of the Hall measurements are provided in supplementary information (figure \ref{Fig.S11}, table \ref{Table S1}). At E$_{\rm IMP}$ = 45~keV, n$_{\rm h}$ increases from 5.77\, x 10$^{15}$~$cm^{-2}$ to 1.66\, x 10$^{16}$~$cm^{-2}$, by raising T$_{\rm DA}$ from 400~$^{\circ}$C to 600~$^{\circ}$C. However, the increase in  T$_{\rm DA}$ also led to the lower [Ga] within the implanted region and loss of superconductivity. The situation is similar for samples with E$_{\rm IMP}$ = 35~keV, where the increase in (n$_{\rm h}$) results in a normal metallic behavior. These results confirm that Ga does not (at least solely) contribute to superconductivity as a dopant, but as a segregated phase (amorphous or crystalline) present within the bulk or at the SiO$_{\rm 2}$/Ge interfaces.

The difference in superconductivity mechanism for PC vs. NC samples was further revealed by 15s of SiO$_{\rm 2}$ etch (using 6:1 buffer oxide etchant(BOE)). Despite no effect on the microstructure (supplementary information figure \ref{Fig.S11}) and minimal reactivity with Ga, the enchant is expected to remove Ga through removal of its host matrix (Ge and SiO$_{\rm 2}$) \cite{johnson_nitrogen_1932}. R$_{\rm s}$(T) measurements illustrated that the BOE etch has a pronounced effect on the superconducting phase at all E$_{\rm IMP}$s. Only the PC sample with E$_{\rm IMP}$ = 80~keV retains its superconductivity after etch, although with lower T$_{\rm c}$ and B$_{\rm c}$ (see supplementary information figure \ref{Fig.S12}). This implies that a portion of the superconductivity for this sample resides sub-surface, possibly induced by Josephson coupling between the Ga-rich clusters that are dispersed within the implanted region \cite{deutscher_upper_1982}. In contrast, in all NC samples with E$_{\rm IMP}$ $\leq$ 45~keV, the zero-resistance state disappeared after etch. At T $\leq$ 1.5~K, the etch effect becomes more dramatic as E$_{\rm IMP}$ becomes smaller, with a fully metallic behavior for the 25~keV sample. Combining the etch results with the TEM and EDS studies, it may be concluded that the superconductivity with zero resistance for NC samples is only provided by the Ga constricted 3--5nm below the SiO$_{\rm 2}$ barrier.

\begin{figure}[ht!]
  \includegraphics[width=0.5\textwidth]{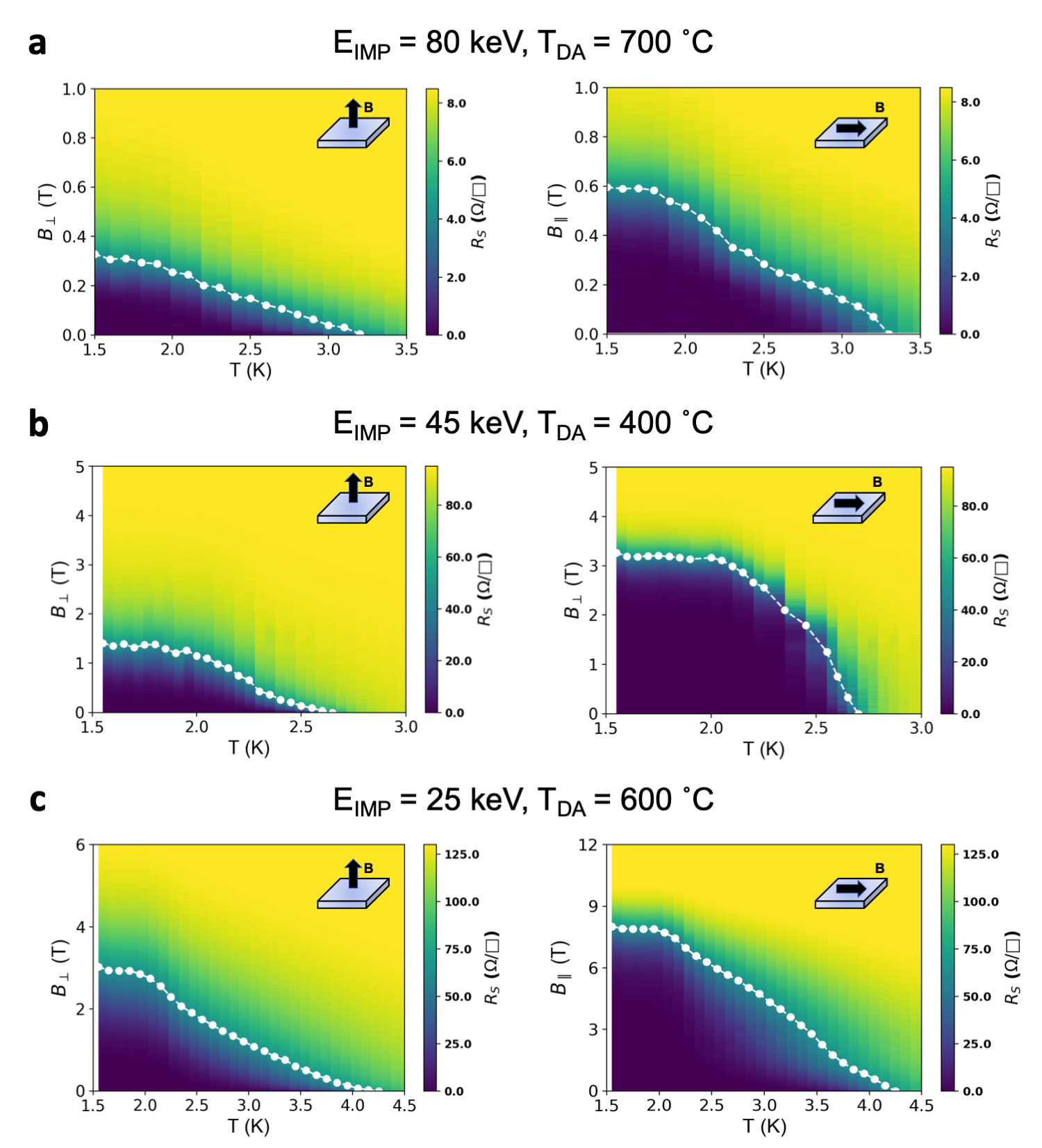}
      \caption{\label{Fig.5} Sheet resistance maps vs. temperature and magnetic field, applied in-plane ($B_{\perp}$) and out-of-plane ($B_{\parallel}$) to the surface of sample. \textbf{a} E$_{\rm IMP}$ = 80 keV \& T$_{\rm DA}$ = 700 $\degree$C. \textbf{b} E$_{\rm IMP}$ = 45 keV \& T$_{\rm DA}$ = 400 $\degree$C. \textbf{c} E$_{\rm IMP}$ = 25 keV \& T$_{\rm DA}$ = 600 $\degree$C. The white overlay plot on each map outlines the normal-superconductor transition boundary, taken at points where R$_{\rm s}$ = 0.5 R$_{\rm s}$.}
\end{figure}

\textbf{Temperature dependence of Resistive Critical Fields.} Next, we turn to study the effect of Ga-doped Ge phase (NC or PC) can have on the  superconductor--normal transitions under magnetic fields. For all superconducting samples labeled as "SC" in figure \ref{Fig.1}c, temperature dependence of R$_{\rm s}$(B) was measured under in-plane (B$_{\rm \perp}$) and out-of-plane (B$_{\rm \parallel}$) magnetic fields as high 12~T. Figure \ref{Fig.5} displays R$_{\rm s}$(B) vs. temperature heat maps for measured for three samples with E$_{\rm IMP}$ = 80~keV (a), 45 keV~(b), and 25 keV~(c). The resistive critical field (B$_{\rm c}$) at each temperature, is defined as the field at which R$_{\rm s}$ = 0.5 $R_{\rm n}$. The overlay plots in figure \ref{Fig.5} indicate temperature dependence of B$_{\rm c}$, which is often known as the boundary for the superconducting-metal phase transition \cite{spivak_theory_2008, fisher_dissipation_1987}. A more comprehensive version of B$_{\rm c}$ phase boundary curves for all "SC" samples is provided in the supplementary information (figure \ref{Fig.S14}). 

\setcitestyle{numbers}

\begin{table}[ht!]
    \centering
    \caption{Summary of superconducting characteristics for three of the Ga-doped Ge samples prepared in this Study}
    \begin{tabular}{|c|c|c|c|c|c|c|}
        \hline
        E$_{\rm IMP}$ & T$_{\rm DA}$  & $T_c$ & $B_{\perp}$ & $B_{\parallel}$ & Clogston & Chandrasekhar\\
         & & & & & limit, Ref.\cite{clogston_upper_1962} & limit, Ref.\cite{chandrasekhar_note_1962} \\
        (keV) & ($\degree$C) & (K) & (T) & (T) & (T) & (T)\\
        \hline
        80 keV & 700 & 3.3 & 0.32 & 0.59 & 5.94 & 8.58  \\
        \hline
        45 keV & 400 & 2.65 & 1.39 & 3.22 & 4.77 & 6.89  \\
        \hline
        25 keV & 600 & 4.25 & 2.97 & 7.95 & 7.65 & 11.05\\
        \hline
    \end{tabular}
    \label{table.1}
\end{table}{}
\setcitestyle{super}

Table \ref{table.1} summarizes the key superconducting parameters extracted from the B$_{\rm c}$--T curves in figure \ref{Fig.5}a--c. While the B$_{\rm c}$(T) for many superconductors follows the Bardeen-Cooper-Schrieffer (BCS) parabolic form B$_{\rm c}$ = B$_{\rm 0}$[1 - (T/T$_{\rm c}$)$^{\rm 2}$], the phase boundaries outlined in figure \ref{Fig.5} show significant deviations from that behavior. Therefore, instead of reporting zero-temperature critical magnetic field (B$_{\rm 0}$) we limit our discussion to B$_{\rm c}$ at 1.55~K, near the base temperature of our cryostat. Table \ref{table.1} also includes the Clogston \cite{clogston_upper_1962} and Chandrasekhar \cite{chandrasekhar_note_1962} upper critical field limits, estimated originally for disordered type-II superconductors with filamentary phases. PC phases with exhibit E$_{\rm IMP}$ = 80~keV have B$_{\rm c}$ values well below these limits in both in-plane and out-of-plane configurations. On the other hand, by going to lower E$_{\rm IMP}$ (e.g. 45~keV \& 25~keV), we saw significant increases in the B$_{\rm c}$ of NC phases, nearing their Clogston limit (see supplementary table \ref{Table S2}). In the extreme case of E$_{\rm IMP}$ = 25~keV, B$_{\parallel}$ even surpassed the limit by 0.3~T. Similar behavior has been reported for thin lead films, where in these studies this is attributed to strong spin-orbit coupling in the 2D metal \cite{nam_ultrathin_2016}.

The in-plane B$_{\rm c}$(T) phase boundary is a macroscopic property that reflects the dimensionality of the superconductive phases involved \cite{matijasevic_films_2001}. The phase boundary behavior is particularly effective in evaluating the dimensionality in layered normal-superconductor systems \cite{krasnov_magnetic_1996, obi_synthetic_2001}. For 3D superconductivity, the in-plane B$_{\rm c}$ is expected to follow the Ginzburg-Landau (GL) linear relationship equation with B$_{\rm c}$ $\propto$ 1 - (T/T$_{\rm c}$). In turn, for 2D superconductivity, the in-plane  B$_{\rm c}$(T) should take the form B$_{\rm c}$ $\propto$ $\sqrt[]{1 - (T/T_{\rm c})}$. However, by a closer look at the in-plane B$_{\rm c}$(T) in figure \ref{Fig.5} and \ref{Fig.S14}, we find that the PC samples exhibit large linear sections that may be connected to a 3D network of coupled Ga superconducting puddles while most NC samples illustrate negative curvatures that could be attributed to a pseudo-2D network that follows B$_{\rm c}$ $\propto$ $\sqrt[]{1 - (T/T_{\rm c})^{n}}$, with $n \geq 1$.

\begin{figure}[ht!]
  \includegraphics[width=0.48\textwidth]{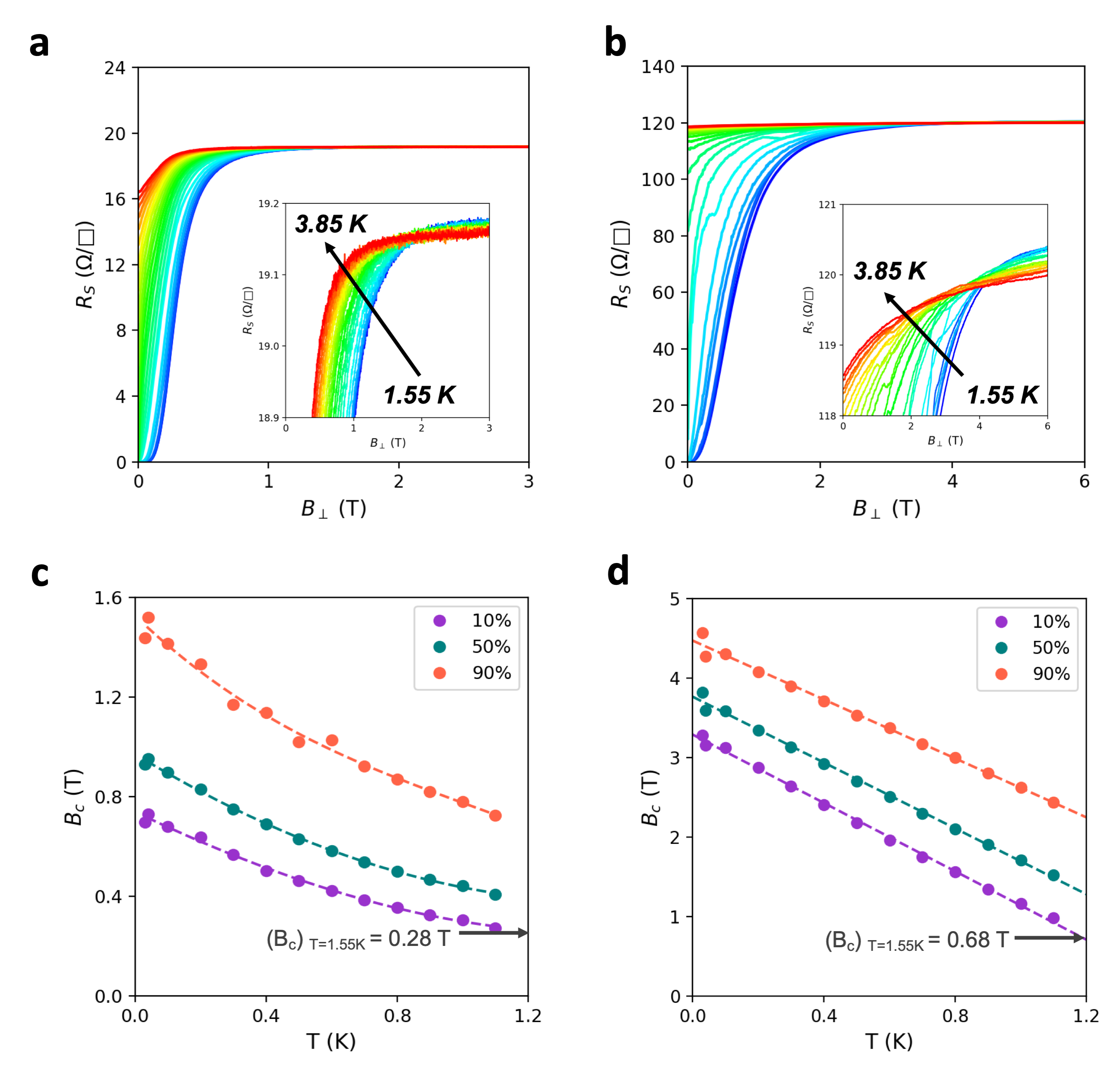}
  \caption{\label{Fig.6} Magnetoresistance and critical field measurements. R$_{\rm s}$ as a function of magnetic field for superconducting Ge samples with \textbf{a} E$_{\rm IMP}$ = 80 keV \& T$_{\rm DA}$ = 700 $\degree$C and \textbf{b} E$_{\rm IMP}$ = 45 keV \& T$_{\rm DA}$ = 400 $\degree$C at 1.55~K--3.85~K. B$_{\rm c}$ vs. temperature extracted from R$_{\rm s}$(B) measurements at 30~mK to 1.1~K for \textbf{c} E$_{\rm IMP}$ = 80 keV \& T$_{\rm DA}$ = 700 $\degree$C and \textbf{d} E$_{\rm IMP}$ = 45 keV \& T$_{\rm DA}$ = 400 $\degree$C. For all those measurements, the magnetic field is applied perpendicular to the sample surface.}
\end{figure}
 
Relatively large B$_{\rm c}$ with complex temperature dependence are not the only signatures of disordered superconductivity observed in the heavily Ga-doped Ge samples. Figure \ref{Fig.6} a \& b display the R$_{\rm s}$(B) isotherms measured at 1.55--3.85 K for two samples with E$_{\rm IMP}$ = 80keV \& T$_{\rm DA}$ = 700 $^{\circ}$C (a) and E$_{\rm IMP}$ = 45keV \& T$_{\rm DA}$ = 400 $^{\circ}$C (b). In both samples crossing points in R$_{\rm s}$(B) were measured at 1.9~T and 4~T, respectively, which are believed to be evidence of quantum phase transition (QPT) in quasi-2D disorder superconductors \cite{xing_quantum_2015,liu_anomalous_2019, zhang_quantum_2019}. R$_{\rm s}$ (T) behavior vs. magnetic field (supplementary figure \ref{Fig.S15}), indicated a more obvious superconductor-metal transition (SMT) from dR/dT$<$0 to dR/dT$>$0 in the sample with E$_{\rm IMP}$ = 45~keV. Because of clear SMT along with better-resolved R$_{\rm s}$(B) crossings, we conducted scaling analysis on this sample to explore the possibility of observing Griffiths singularity behavior with divergent product of correlation length exponent ($\nu$) and dynamical critical exponent ($z$) \cite{markovic_thickness--magnetic_1998, fisher_boson_1989}. Details of  scaling analysis are provided in the supplementary materials (see figure \ref{Fig.S16}). For the sample with E$_{\rm IMP}$ = 45~keV, scaling analysis yielded $z\nu$ = 2.58 $\pm$ 0.46 at T = 1.55--1.95~K range, followed by $z\nu$ = 0.29 $\pm$ 0.01 at T = 2.15--2.55~K range. Similar analysis on a sample with E$_{\rm IMP}$ = 35~keV led to $z\nu$ of 0.65 $\pm$ 0.04 at T = 1.55--1.95~K and 0.4 $\pm$ 0.01 at T = 2.15--2.55~K (supplemental figure \ref{Fig.S17}).  While these values do not establish a trend to a divergent dynamical critical exponent, the general $z\nu$ behavior may warrant further investigation into their SMT at near-zero temperatures and high magnetic fields.

Another signature of anomaly was observed in these samples when B$_{\rm c}$ temperature dependence was measure at near-zero temperatures (i.e. 35~mK--1.1~K). As shown in \ref{Fig.6} c \& d, both samples with E$_{\rm IMP}$ = 80 keV \& T$_{\rm DA}$ = 700 $^{\circ}$C (c) and E$_{\rm IMP}$ = 45 keV \& T$_{\rm DA}$ = 400 $^{\circ}$C (d), show anomalous rise in $B_{\rm c}$ as temperature approaches 0~K . This anomaly persists regardless of the definition used for $B_{\rm c}$, including at critical sheet resistance $R_{\rm c}$ = 0.1 $R_{\rm n}$, 0.5 $R_{\rm n}$ and 0.9 $R_{\rm n}$. This $B_{\rm c}$ vs. T upturn is yet another evidence of disorder in these system. The more interesting feature is the difference between the $B_{\rm c}$ upturn between the two samples; from exponential in figure \ref{Fig.6}c to linear in figure \ref{Fig.6}d. The exponential behavior can be explained by a model of superconducting island weakly coupled via Josephson effect, in which the value of the B$_{\rm c}$ is determined by an interplay between proximity effect and quantum phase fluctuations \cite{galitski_disorder_2001}. In turn, the linear B$_{\rm c}$(T) anomaly has been recently attributed to vortex glass ground states and their thermal fluctuations present in disordered thin-films \cite{sacepe_low-temperature_2019}.\\ 

\noindent
\textbf{Discussion}

In this work, we demonstrated pathways to control the structural and physical characteristics of superconducting phases prepared by hyperdoping of Ge via Ga$^{\rm +}$ implantation. The parameter space for observing superconductivity in hyperdoped Ge has been clearly expanded to temperatures far below Ge melting point (as low as 350 $^{\circ}$C) and implantation energies as low as 25~keV. The phase of the superconductor may be tuned  by the implantation energy and dopant activation temperatures. Two distinct phases of superconducting materials are recognized: i) 3D poly-crystalline (e.g. at E$_{\rm IMP}$80~keV, T$_{\rm DA}$ = 700 $^{\circ}$C ) which is consistent with earlier studies; ii) a new 2D nano-crystalline phase (e.g. at E$_{\rm IMP}$45~keV, T$_{\rm DA}$ = 400 $^{\circ}$C ). We find signatures that the nano-crystalline phase superconductivity stems from few $nm$ thick Ga layer accumulated near the top surface, leading to systems with high quasi-2D disorder as evidenced by various anomalies in critical field behaviors. Both phases showed signatures of strong disorder despite the difference in underlying superconductivity mechanisms. This warrants further investigations into tunability of disorder in these material systems as test-beds for quantum phase transition studies as well as platforms for superconducting circuits with high kinetic inductance.\\

\noindent
\textbf{Methods}

\noindent
\textbf{Fabrication.} Undoped Ge(100) wafers grown by floating zone method with room-temperature resistivity of 40 $\Omega.cm$ were used. Prior to implantation, Ge native oxide was etched cyclic immersion in 10\% HF solution and DI--H$_{\rm 2}$O. This was followed by the deposition of 30~$nm$ thick SiO$_2$ top barriers via plasma-enhanced chemical vapor deposition (PECVD). The oxide barrier helps minimize direct damage to the Ge substrates during the ion implantation process. Ga$^{\rm +}$ ion implantation processes (by Kroko Inc.) were carried out at 25~keV, 35~keV, 45~keV, and 80~keV with a fixed ion fluence of 4e16 cm$^{\rm -2}$. Throughout the implantation substrates were held at room temperature. After implantation dopants were activated using rapid thermal annealing (RTA) under 5 standard liter per minute (SLPM) of N$_{\rm 2}$ flow and at 350~$^{\circ}$C to 800~$^{\circ}$C, for 1~min.\\

\noindent
\textbf{Transport Measurements.} Electrical properties of the samples were evaluated by measuring the differential resistance (i.e. $dV/dI$), using lock-in amplifiers, for 5 $mm$ x 5 $mm$ square samples in Van der Pauw (VdP) geometry. The AC excitation current for the measurements varied between 1 and 20 $\mu$A. Measurements down to 1.5~K were performed in a Teslatron PT (Oxford Instruments) cryogen-free refrigerator with maximum magnetic field of 12~T (along z-axis). Measurements below 1.2~K were carried out in a Triton dilution refrigerator (Oxford Instruments) with a 3-axis vector magnet, and maximum z field of 6T.  Hall measurements were performed on L-shaped bars. Hall bars were fabricated by UV photolithography followed by reactive ion etching of the mesa using CF$_{\rm 4}$/O$_{\rm 2}$ gas mixtures for 2-5 min. The resulting mesa heights varied between 500 $nm$ and 1.2 $\mu m$ (further details in supplementary information).\\

\noindent
\textbf{Structural and Chemical Characterization.} Micro-Raman spectroscopy was performed using a Horiba Xplora $\mu$-Raman system with a 532 $nm$ laser and an objective lens at 1000x magnification. Time-of-Flight Secondary ion mass spectrometer (ToF-SIMS) measurements were performed using Bi$^{\rm +}$ and Ga$^{\rm +}$ ions for detection and Ar$^{\rm +}$ ions for milling. Ion detection was done in positive ion mode by monitoring $^{28}$Si, $^{72}$Ge, $^{74}$Ge, $^{69}$Ga and $^{12}$C. Atomic force microscopy (AFM) was performed in order to determine the surface morphology of the superconducting films in details. A Bruker Dimension Fastscan scanning probe microscopy system was used in tapping mode (ScanAsyst mode). The AFM probes used for the measurements were Bruker FASTSCAN-B, made of silicon nitride with triangular tips of 5--12 $nm$ radius.\\

\noindent
\textbf{Electron Microscopy.} The crystal morphology of the structures were examined with with transmission electron microscopy (TEM) in a JEOL ARM200F equipped, with a spherical aberration corrector for probe mode, operated at 200~keV. The composition of each films was studied with energy dispersive spectroscopy (EDS). The samples were prepared with cross-sectional tripod polishing to ~20 $\mu m$ thickness, followed by shallow angle Ar$^{\rm +}$ ion milling with low beam energies ( $\leq$ 3~keV),  and LN$_{\rm 2}$ stage cooling in a PIPS II ion mill.\\

\bibliographystyle{naturemag}
\makeatletter \renewcommand\@biblabel[1]{#1.} \makeatother
\bibliography{Ge_SC_refs}

\vspace{\baselineskip}

\textbf{Data Availability.}

The data that support the findings of this study are available within the paper and its supplementary Information. Additional data are available from the corresponding author upon request.\\


\textbf{Acknowledgements}

 The authors thank Emanuel Tutuc and Fatemeh Barati for fruitful discussions. NYU team is supported by AFOSR Grant No. FA9550-16-1-0348 and ARO Grant No. W911NF1810115 . K.S. thanks Davood Shahjerdi and Edoardo Cuniberto for their assistance with the Raman spectroscopy measurements. K.S. and T.N. thank Tai-De Li of CUNY ASRC surface science suite is acknowledged for his assistance with ToF-SIMS measurements. J.Y. acknowledges funding from the ARO/LPS QuaCGR fellowship reference W911NF1810067.\\

\textbf{Authors contributions.}

K.S. and T.D.N. synthesized the superconducting film structures and fabricated micro-devices.  K.S., T.N., W.L.S., A.L.F, M.H., J.Y. and W.M. performed the measurements with J.S. providing input. K.S., T.N. and M.C.D. performed data analysis. K.S. and J.S. conceived the experiments. All authors contributed to interpreting the data. The manuscript was written by K.S., W.L.S, M.C.D., J.Y. and J.S. with suggestions from all other authors.\\

\renewcommand{\thefigure}{\textbf{S\arabic{figure}}}
\renewcommand{\theequation}{\textbf{S\arabic{equation}}}
\setcounter{figure}{0}
\setcounter{equation}{0} 

\clearpage
\newpage
\onecolumngrid

\section{Supplementary information}

\renewcommand{\thefigure}{S\arabic{figure}}
\setcounter{figure}{0}

\begin{figure}[ht!]
    \centering
    \includegraphics[width=0.8\textwidth]{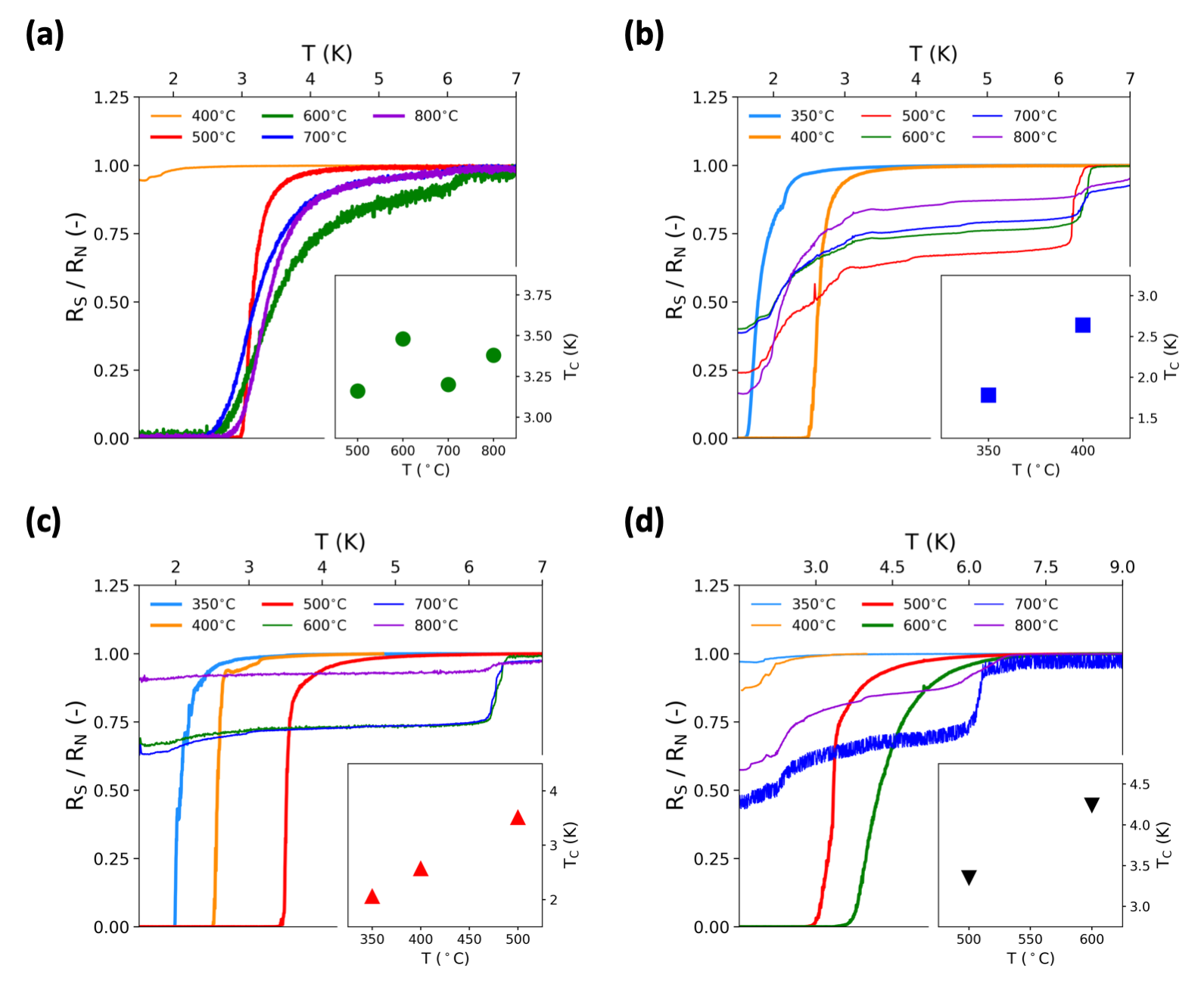}
    \caption{Normalized resistance vs. temperature for the majority of the samples tested in this study with implantation energies of (a) 80 keV, (b) 45keV (b), (c) 35keV,  and (d) 25 keV. The inset for each plot displays the variations of T$_{\rm c}$, at a fixed E$_{\rm {IMP}}$, as a function of annealing temperature.}
    \label{Fig.S1}
\end{figure}

\begin{figure}[ht!]
  \includegraphics[width=0.85\textwidth]{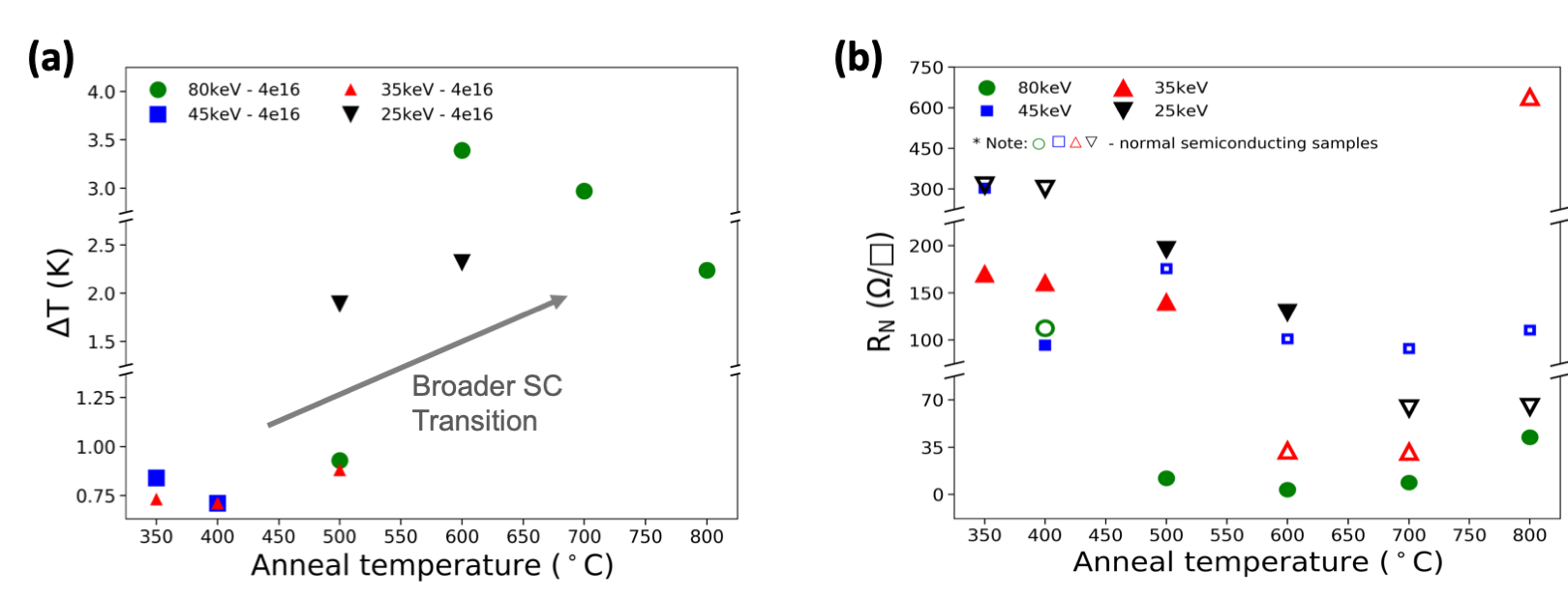}
  \caption{Transition width ($\Delta$T) and normal resistance (R$_N$) distribution for the Ga-implanted Ge samples: (a) transition width for all 10 superconducting samples, defined by the temperature difference between points with 0.95R$_N$ and 0.05 R$_N$; (b) R$_N$ for all samples (normal and superconducting) as a function of anneal temperature.}
  \label{Fig.S2}
\end{figure}

\begin{figure}[ht!]
  \includegraphics[width=0.9\textwidth]{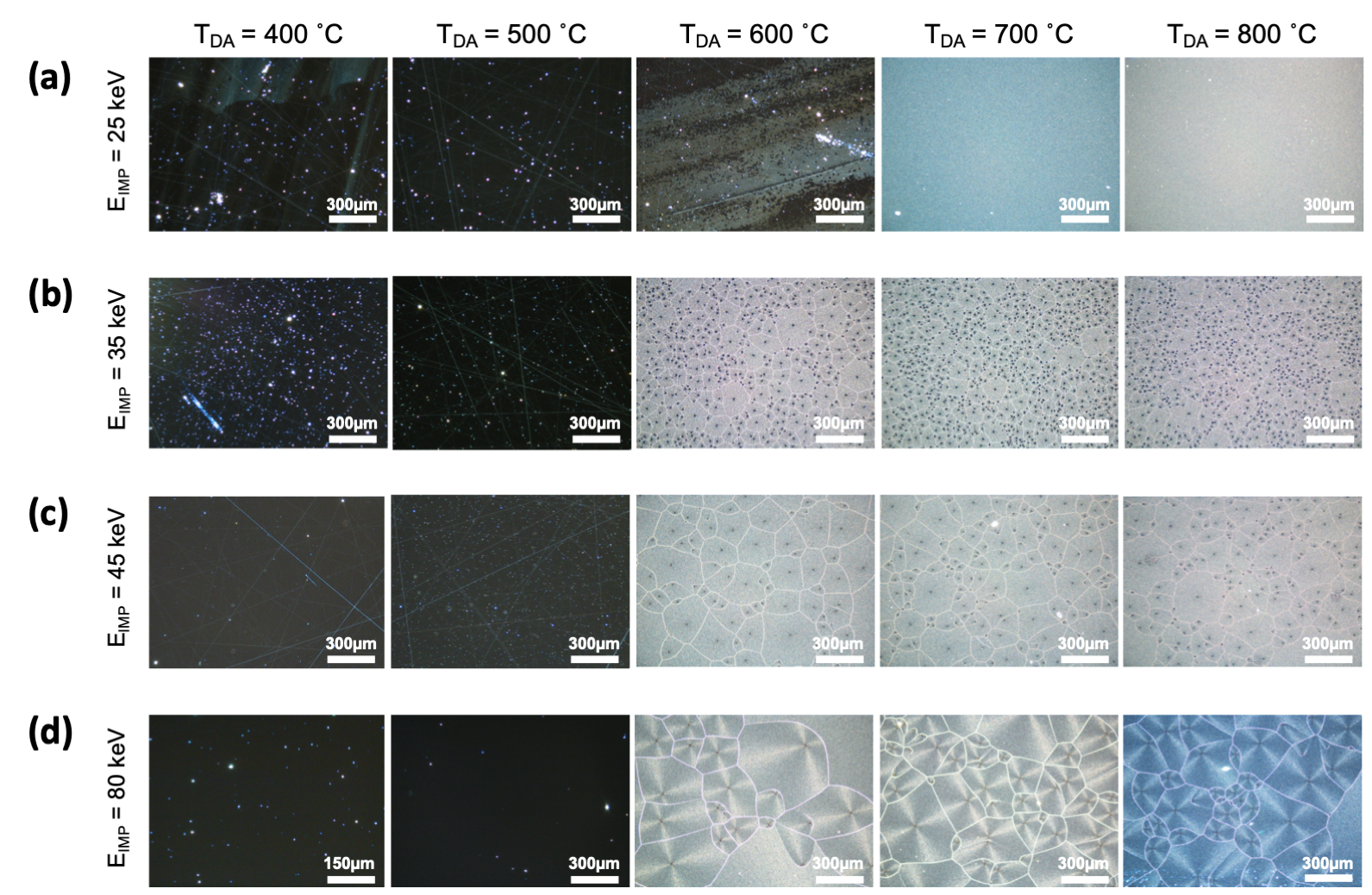}
  \caption{Optical microscopy images for Ga-implanted Ge samples with implantation energy of (a) 25keV, (b) 35keV, (c) 45keV and, (d) 80keV, after activation annealing at 400-800 $^\circ$C.}
  \label{Fig.S3}
\end{figure}

\begin{figure}[ht!]
  \includegraphics[width=0.9\textwidth]{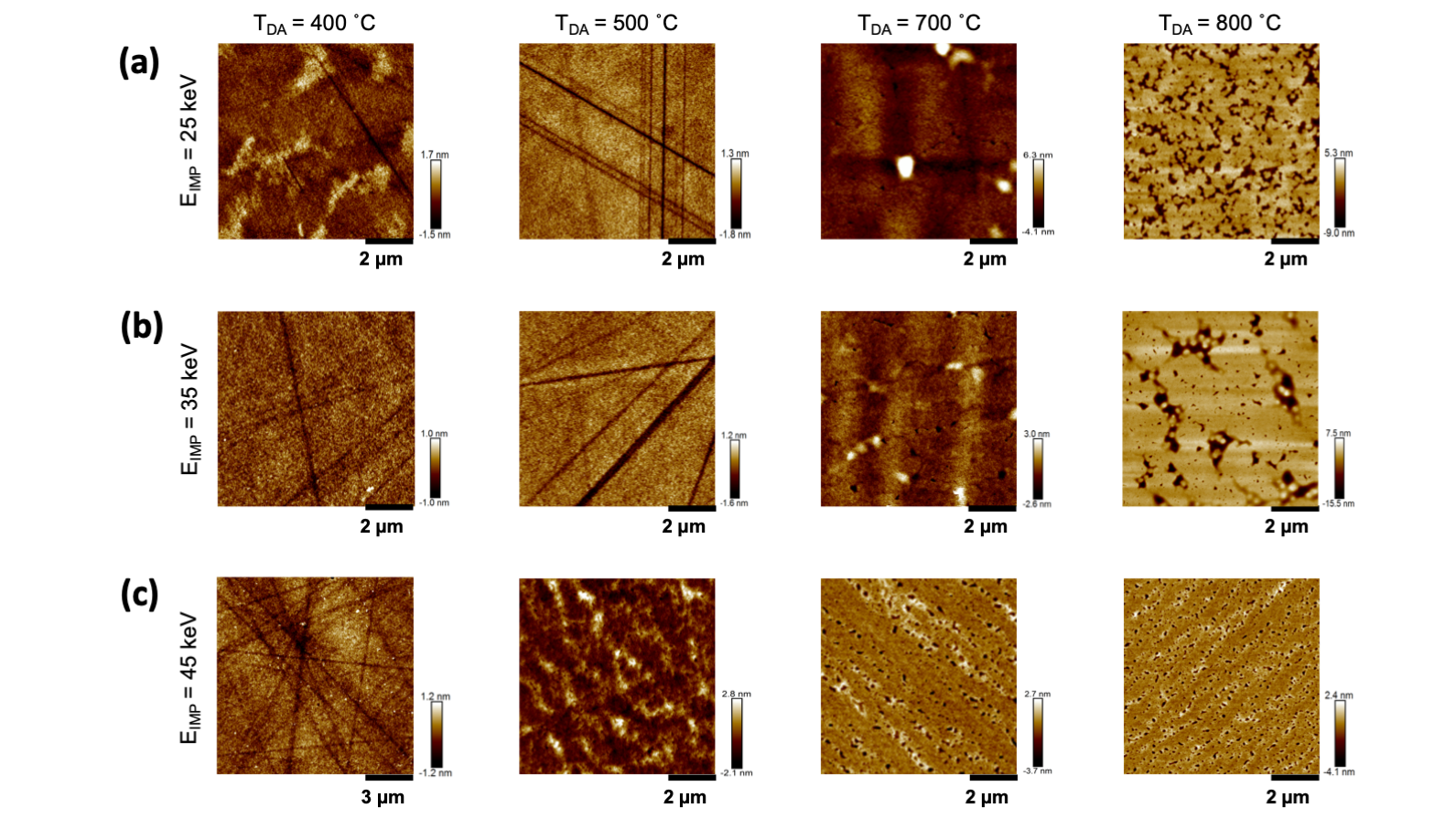}
  \caption{Surface topographical maps for Ga-implanted Ge samples with implantation energy of (a) 25keV, (b) 35keV and, (c) 45keV, after activation annealing at 400-800 $^\circ$C. }
  \label{Fig.S4}
\end{figure}

\begin{figure}[ht!]
  \includegraphics[width=350pt, height = 112pt]{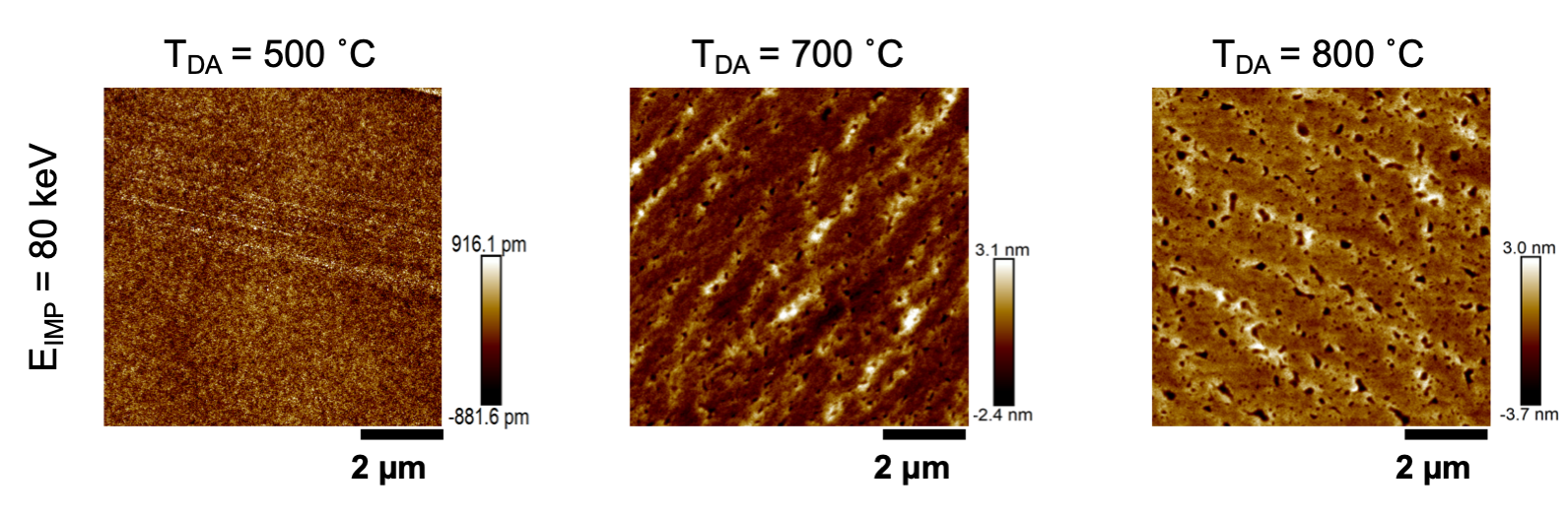}
  \caption{Surface topographical maps for Ga-implanted Ge samples with implantation energy of 80keV after activation annealing at 500-800 $^\circ$C. }
  \label{Fig.S5}
\end{figure}

\begin{figure}[ht!]
  \includegraphics[width=0.95\textwidth]{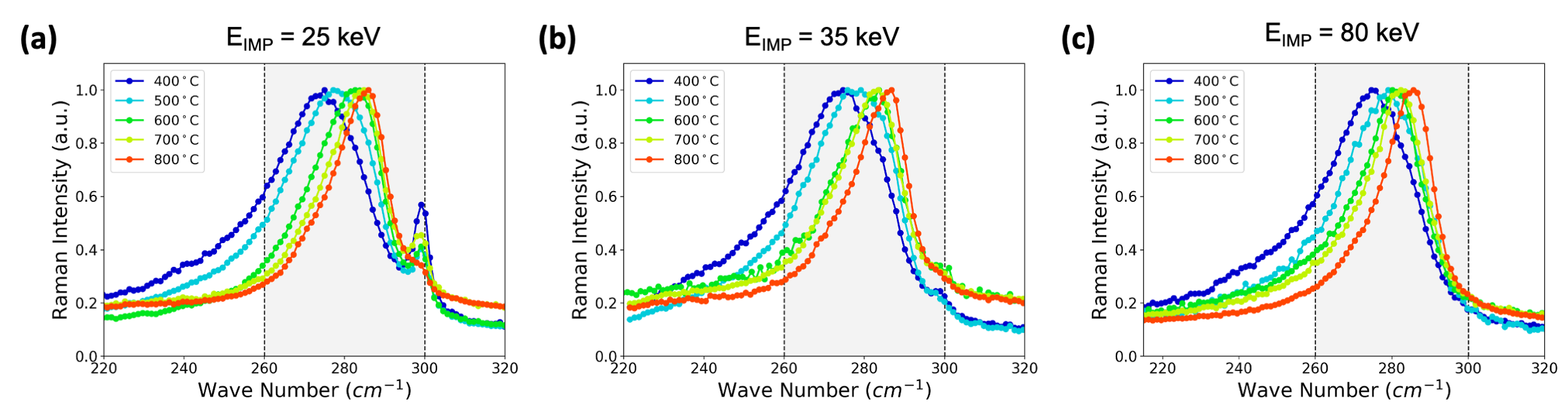}
  \caption{Raman spectrum as a function of activation annealing temperature for Ga-implanted Ge samples with implantation energy of (a) 25keV, (b) 35keV and, (c) 80keV. The annealing temperature range included is 400-800 $^\circ$C. }
  \label{Fig.S6}
\end{figure}

\begin{figure}[hb!]
  \includegraphics[width=0.75\textwidth]{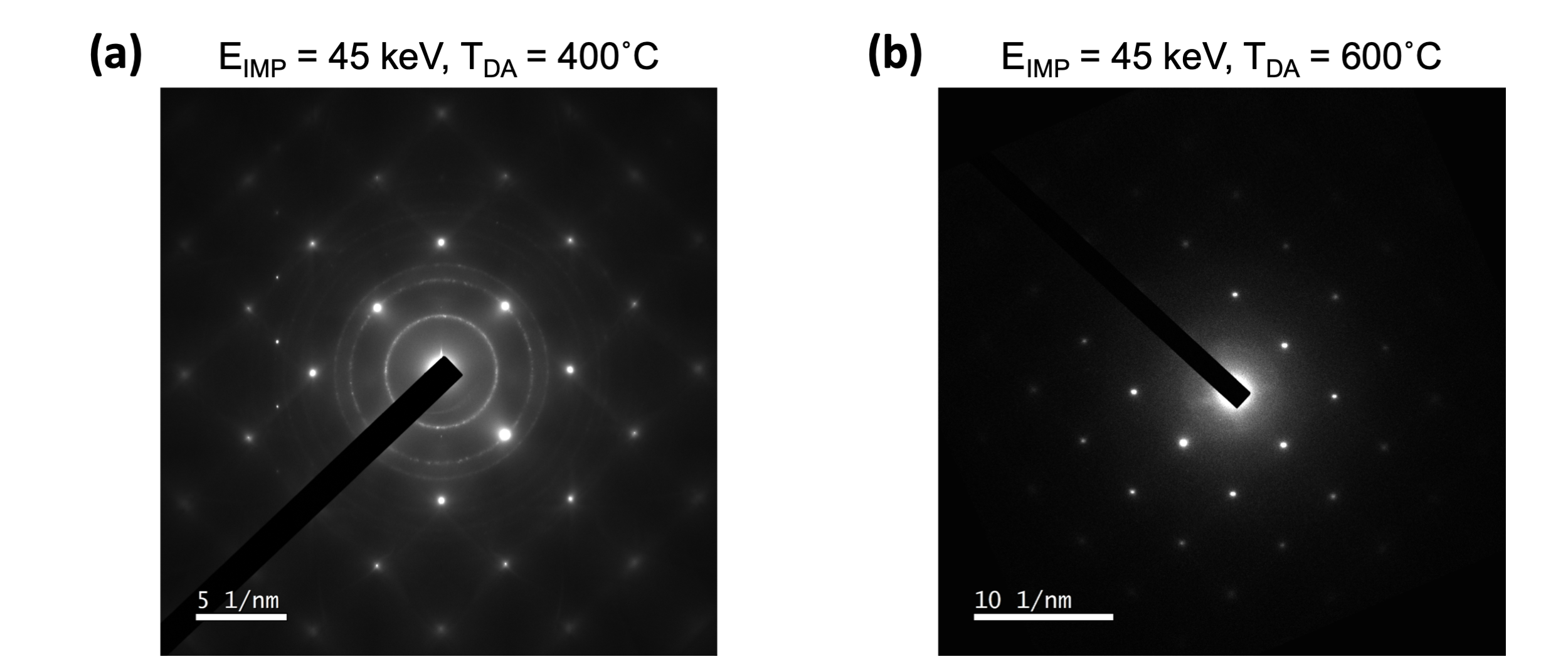}
  \caption{Electron diffraction patterns taken via TEM on two Ge samples with (a) E$_{\rm IMP}$ = 45 keV \& T$_{\rm DA}$ = 400 $\degree$C; (b) E$_{\rm IMP}$ = 45 keV \& T$_{\rm DA}$ = 600 $\degree$C. For the sample with lower anneal temperature, the observed rings signify the presence of a nanocrystalline phase in addition to single-crystalline Ge substrates.}
  \label{Fig.S7}
\end{figure}

\begin{figure}[ht!]
  \includegraphics[width=0.85\textwidth]{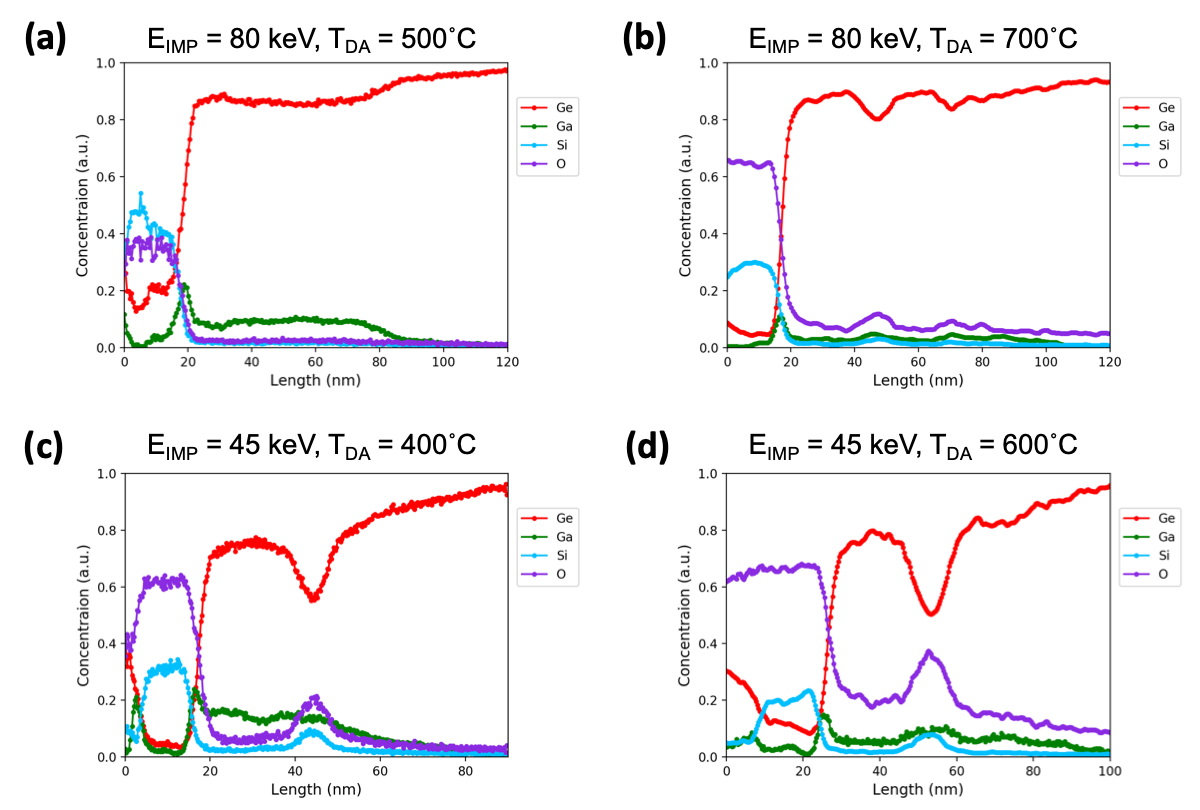}
  \caption{EDS elemental distribution line scans for Ge, Ga, Si and O for four hyperdoped Ge samples including: a)E$_{\rm IMP}$ = 80 keV \& T$_{\rm DA}$ = 500 $\degree$C; (b) E$_{\rm IMP}$ = 80 keV \& T$_{\rm DA}$ = 700 $\degree$C; (c) E$_{\rm IMP}$ = 45 keV \& T$_{\rm DA}$ = 400 $\degree$C; (d) E$_{\rm IMP}$ = 45 keV \& T$_{\rm DA}$ = 600 $\degree$C. The line traces clearly show how higher anneal temperature would deplete the film from Ga. Additionally, at lower E$_{\rm IMP}$ a band if $SiO_x$ cluster can be easily observed as a result of $SiO_2$ cap recoil into the substrate.}
  \label{Fig.S8}
\end{figure}

\begin{figure}[ht!]
  \includegraphics[width=0.85\textwidth]{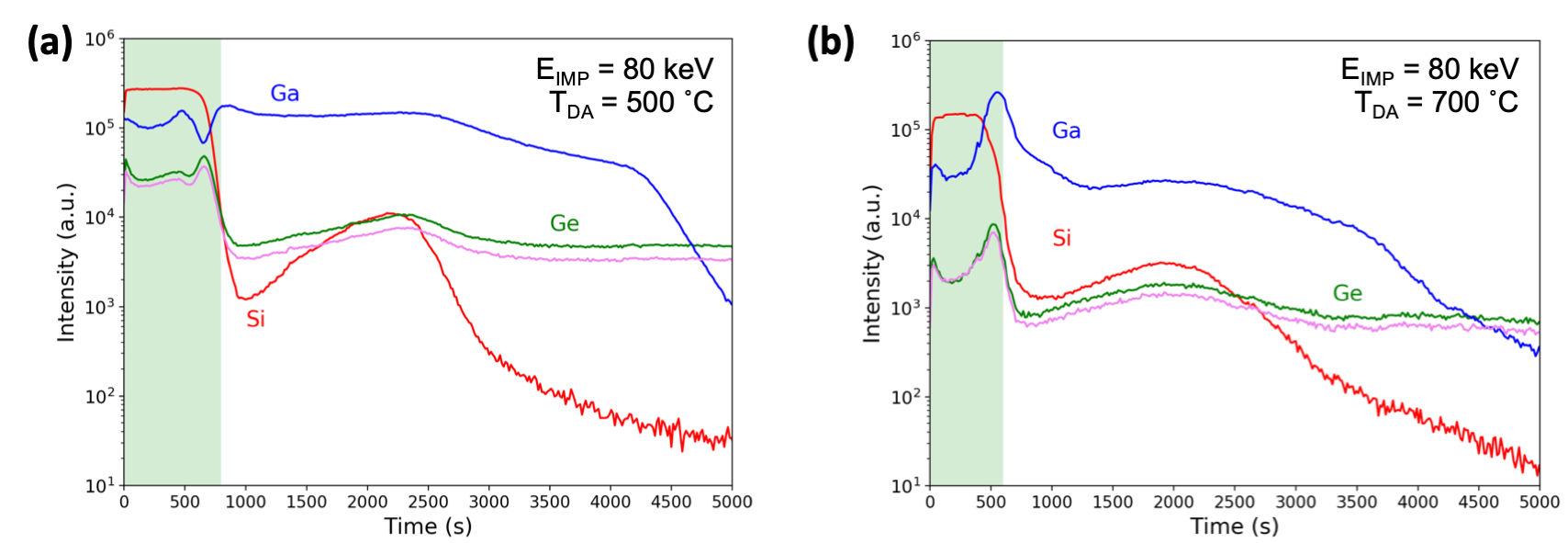}
  \caption{Depth profiles for $^{28}$Si, $^{72}$Ga, $^{72}$Ge and $^{74}$Ge on samples with  E$_{\rm IMP}$ = 80 keV that underwent activation annealing at T$_{\rm DA}$ = 500 $\degree$C (a) and T$_{\rm DA}$ = 700 $\degree$C (b). Depth profiling was carried out by by time-of-flight secondary ion mass spectrometer (ToF-SIMS) with Ar$^+$ as the sputtering ion and Bi$^+$ as the analysis ion.}
  \label{Fig.S9}
\end{figure}

\clearpage

\section*{Hall Measurements}

In order to determine the hole concentration in hyper Ga-doped Ge samples, Hall measurements were performed on L-shaped Hall bars as shown in figure \ref{Fig.S13}a. The  $\sim$ 150~$\mu$m wide Hall bars had two contacts, one at each end, for current injection, as well as four pairs of contacts for Hall voltage measurements in two orthonormal directions. Measurements were carried out by sweeping the fields (B) from -6~T to 6~T, perpendicular to the sample surface, while measuring Hall coefficients (R$_H$) in two directions as show in figure \ref{Fig.S13}b. Typically, R$_H$ vs. B were recorded from 3~K to 200~K. However, the linear behavior corresponding to hole carriers was only observed below 30~K. This is due to the fact that the Ge substrates have an equilibrium Sb concentration of 1e14 cm$^{-2}$, making them lightly n-type. The p-n junction that forms between between the Ga-doped region and the substrate is only fully depleted at temperatures below 30~K. Figure \ref{Fig.S13}c displays the R$_H$(B) behavior for a Ge sample with E$_{\rm IMP}$ = 45keV \& T$_{\rm DA}$ = 400 $^{\circ}$C. The negative slope confirms the presence of holes as majority carriers in the sample. All eight samples that were subjects of the Hall measurements showed a negative slope. From the magnitude of the slope ($m$) we determined the hole density ($n_h$) in cm$^{-2}$ using the following equation:

\begin{equation}
    n_h = \frac{10^{-4}}{m.q}
    \label{EqCarriers}
\end{equation}

where $q$ is the elementary charge of holes (1.602e-19 C). Since no significant shift was observed in the R$_H$(B) slope vs. temperature, only one value is reported as $n_h$ for each sample. Table \ref{Table S1} summarizes the processing conditions and $n_h$ for all 80 samples. In addition to the measured hole density, Ga density in Ge --estimated by SRIM Monte Carlo simulation-- is included in the last column.


\begin{figure}[ht!]
  \includegraphics[width=\textwidth]{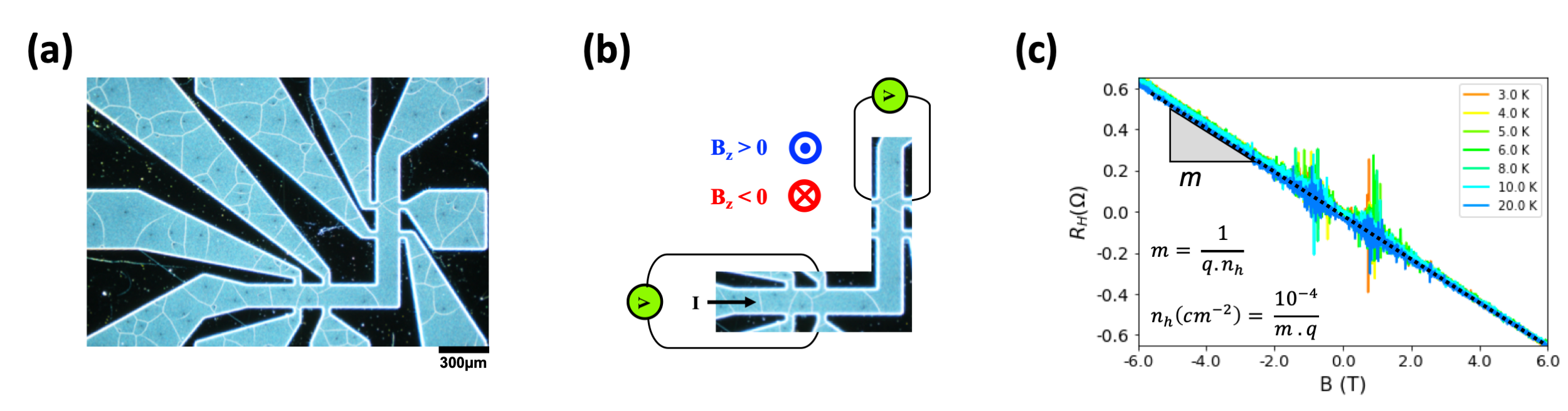}
  \caption{Details of the carrier density measurements in hyperdoped Ge samples prepared for this study including: (a) An optical micrograph of the L-shaped bars fabricated for Hall measurements via photolithography and reactive ion etching; (b) The measurement configuration for the Hall resistance with the magnetic filed perpendicular to the sample surface; (c) An example of the measured Hall coefficient vs. magnetic field measured on a Ge sample with  E$_{\rm IMP}$ = 45keV \& T$_{\rm DA}$ = 400 $^{\circ}$C, within the typical temperature range of 3~K--20~K.}
  \label{Fig.S10}
\end{figure}


\renewcommand{\thetable}{S\arabic{table}}
\setcounter{table}{0} 
\newcolumntype{C}[1]{>{\centering\arraybackslash}p{#1}}

\begin{ThreePartTable}
    \begin{TableNotes}
        \item[a] Estimated hole density by TRIM simulations.
        \item[b] Sample is not superconducting.
    \end{TableNotes}
    \begin{longtable}{C{2cm}C{2cm}C{2cm}C{2cm}C{2cm}}
        \caption{Summary of hole density for eight Ga-doped Ge samples estimated by Hall measurements on L-shpaed bars.}\\
        \hline
        E$_{\rm IMP}$ & T$_{\rm DA}$  & $T_c$ & $n_h$ & $n_{Ga}$ \tnote{a}\\
        (keV) & ($\degree$C) & (K) & ($cm^{-2}$) & ($cm^{-2}$)\\
        \hline
        \endhead
        \cmidrule{2-2}
        \multicolumn{2}{r}{\textit{continued}}
        \endfoot
        \bottomrule
        \insertTableNotes
        \endlastfoot
        80 keV & 800 & 3.4 & 1.07\, x 10$^{16}$ & 3.46\, x 10$^{16}$ \\
        80 keV & 500 & 3.0 & 5.48\, x 10$^{15}$ \\
        \hline
        45 keV & 600 & N/A\tnote{b} & 1.66\, x 10$^{16}$ & 1.90\, x 10$^{16}$ \\
        45 keV & 400 & 2.6 & 5.77\, x 10$^{15}$ \\
        \hline
        35 keV & 600 & N/A\tnote{b} & 1.42\, x 10$^{16}$ & 1.04\, x 10$^{16}$ \\
        35 keV & 400 & 2.6 & 5.70\, x 10$^{15}$ \\
        \hline
        25 keV & 700 & N/A\tnote{b} & 1.90\, x 10$^{15}$ & 2.38\, x 10$^{15}$ \\
        25 keV & 500 & 3.3 & 3.41\, x 10$^{15}$ \\
        \hline
        \label{Table S1}
    \end{longtable}
\end{ThreePartTable}


\begin{figure}[ht!]
  \includegraphics[width=0.8\textwidth]{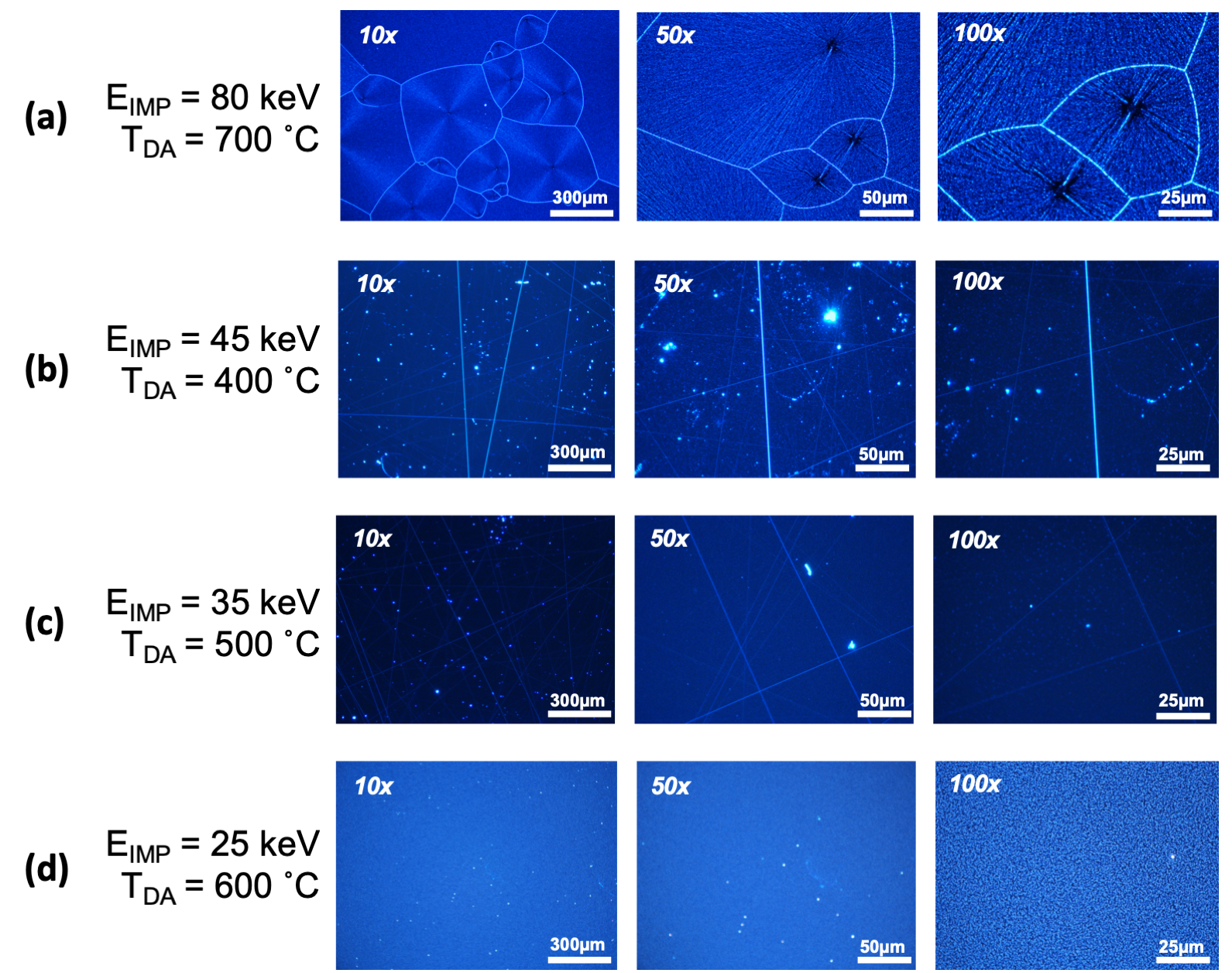}
  \caption{Optical microscopy images of four superconducting samples etched by 15s of immersion in 6:1 buffer oxide etchant (BOE) followed by 30s of DI $H_2O$ rinse, including: (a)E$_{\rm IMP}$ = 80 keV \& T$_{\rm DA}$ = 700 $\degree$C; (b) E$_{\rm IMP}$ = 45 keV \& T$_{\rm DA}$ = 400 $\degree$C; (c) E$_{\rm IMP}$ = 35 keV \& T$_{\rm DA}$ = 500 $\degree$C; (d) E$_{\rm IMP}$ = 25 keV \& T$_{\rm DA}$ = 600 $\degree$C.}
  \label{Fig.S11}
\end{figure}

\begin{figure}[ht!]
  \includegraphics[width=0.7\textwidth]{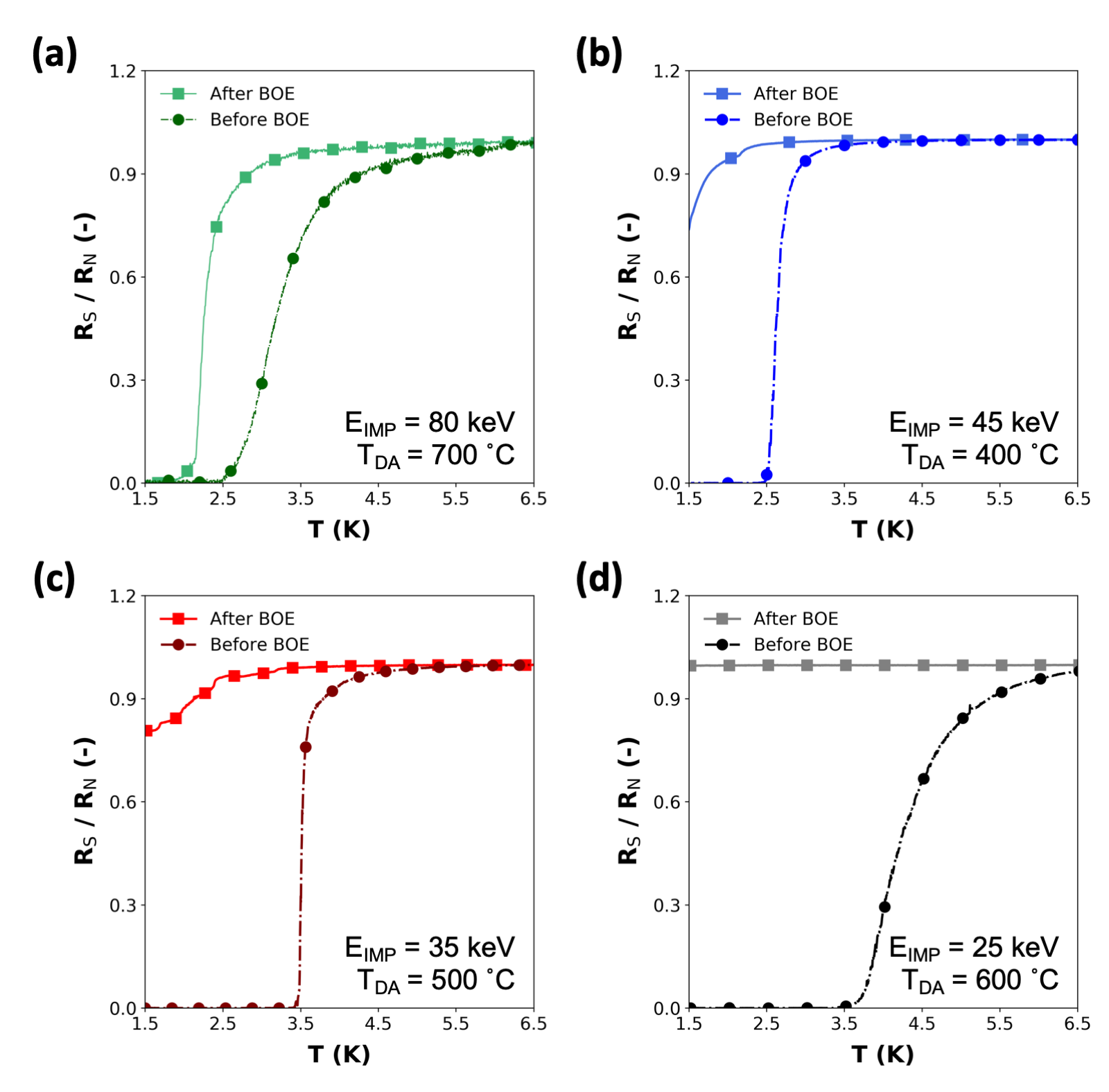}
  \caption{Normalized resistance vs. temperature for four samples before (black solid line) and after etching the $SiO_2$ cap (red dotted line) including: (a)E$_{\rm IMP}$ = 80 keV \& T$_{\rm DA}$ = 700 $\degree$C; (b) E$_{\rm IMP}$ = 45 keV \& T$_{\rm DA}$ = 400 $\degree$C; (c) E$_{\rm IMP}$ = 35 keV \& T$_{\rm DA}$ = 500 $\degree$C; (d) E$_{\rm IMP}$ = 25 keV \& T$_{\rm DA}$ = 600 $\degree$C. For oxide etch, 15s dip in 6:1 buffer oxide etchant (BOE) followed by 30s of DI $H_2O$ was employed.}
  \label{Fig.S12}
\end{figure}


\begin{figure}[ht!]
  \includegraphics[width=0.8\textwidth]{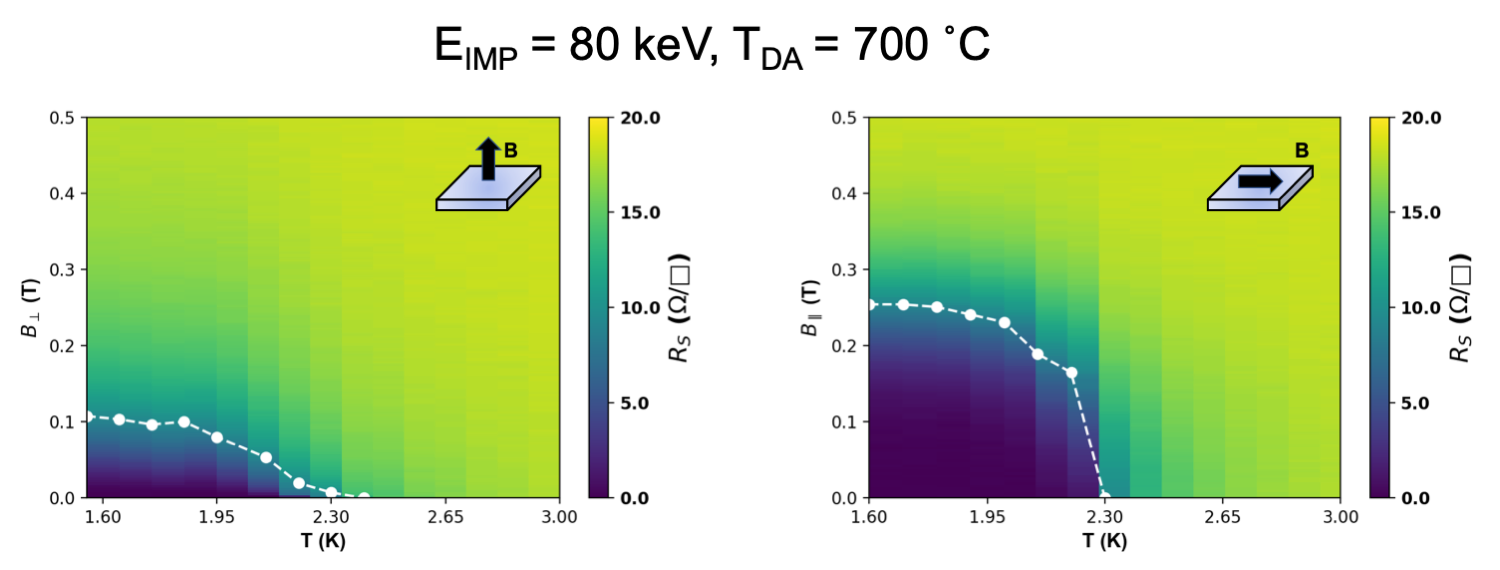}
  \caption{Sheet resistance as a function of temperature and magnetic field, both perpendicular ($B_{\perp}$) and parallel ($B_{\parallel}$) to the surface of the Ge sample with E$_{\rm IMP}$ = 80 keV \& T$_{\rm DA}$ = 700 $\degree$C. This sample was subject to oxide barrier removal by BOE. The white lines on each map display the Normal-Superconductor phase transition boundary, assumed as the point where resistance is equal to 50\% of the $R_N$.}
  \label{Fig.S13}
\end{figure}

\clearpage

\begin{figure}[ht!]
  \includegraphics[width=\textwidth]{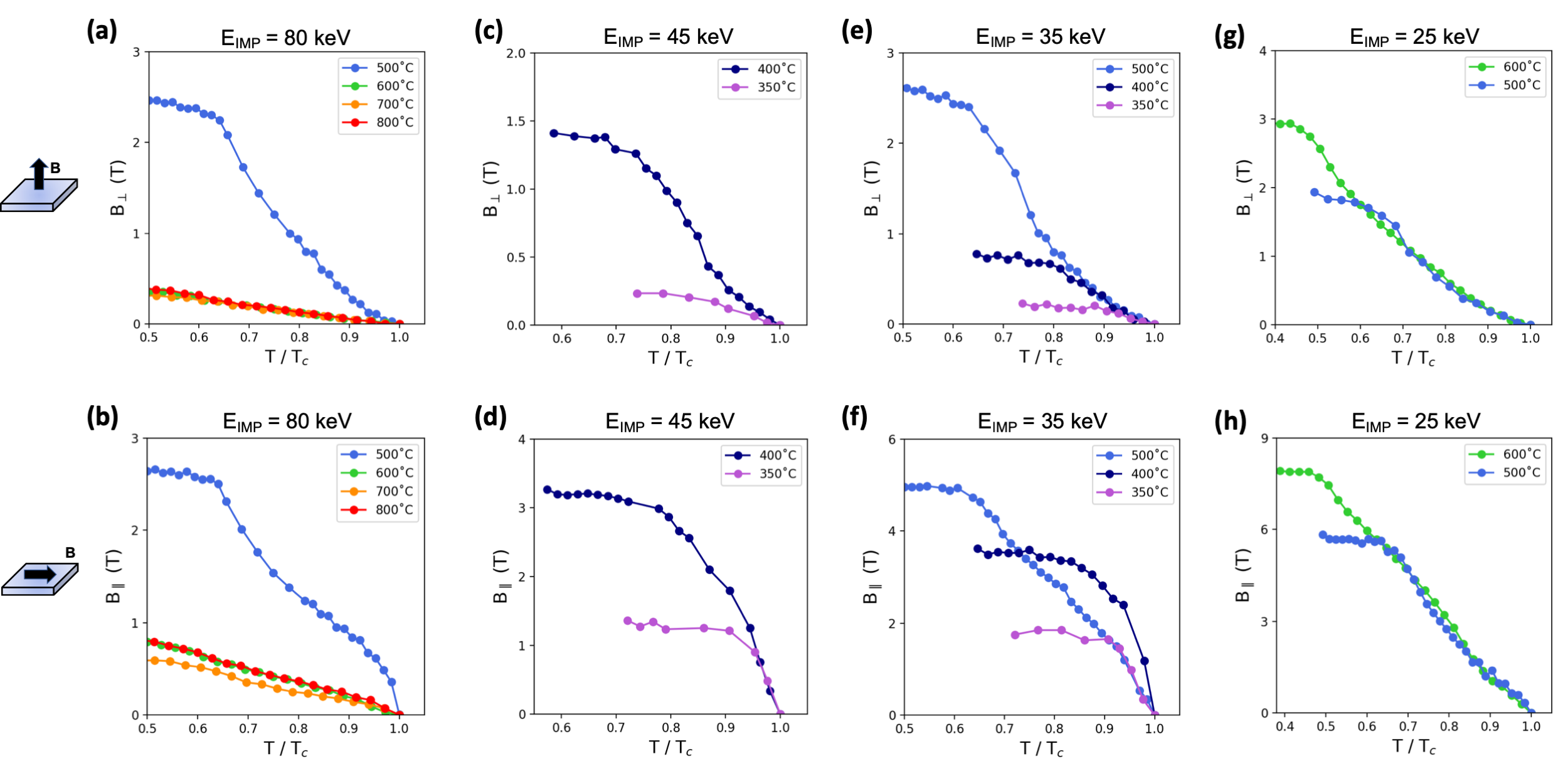}
  \caption{An overview of the B$_c$(T) behavior for all the superconducting samples in out-of-plane and in-plane magnetic fields with implantation energies of 80 keV (a, b), 45 keV (c, d), 35 keV (e, f), and 25 keV (g, h).}
  \label{Fig.S14}
\end{figure}

\begin{ThreePartTable}
    \begin{TableNotes}
        \item[a] Averaged over the temperature range of 1.5--1.6K.
        \item[b] Clogston limit is B$_{c} \sim$ 1.8 T$_c$.
        \item[c] Chandrasekhar limit is B$_{c} \sim$ 2.6 T$_c$.
    \end{TableNotes}
    \begin{longtable}{C{1.5cm}C{1.5cm}C{1.5cm}C{1.5cm}C{1.5cm}C{2cm}C{2.5cm}}
        \caption{Summary of superconducting parameters for all samples that exhibited the zero-resistance state.}\\
        \hline
        E$_{\rm IMP}$ & T$_{\rm DA}$  & $T_c$ & $B_{\perp}$\tnote{a} & $B_{\parallel}$\tnote{a} & Clogston\tnote{b} & Chandrasekhar \tnote{c}\\
        (keV) & ($\degree$C) & (K) & (T) & (T) & (T) & (T)\\
        \hline
        \endhead
        \cmidrule{2-2}
        \multicolumn{2}{r}{\textit{continued}}
        \endfoot
        \bottomrule
        \insertTableNotes
        \endlastfoot
        80 keV & 800 & 3.50 & 0.40 & 0.81 & 6.30 & 9.10\\
        80 keV & 700 & 3.3 & 0.32 & 0.59 & 5.94 & 8.58\\
        80 keV & 600 & 3.60 & 0.37 & 0.81 & 6.48 & 9.36\\
        80 keV & 500 & 3.20 & 2.48 & 2.65 & 5.76 & 8.32\\
        \hline
        45 keV & 400 & 2.65 & 1.39 & 3.22 & 4.77 & 6.89\\
        45 keV & 350 & 2.10 & 0.22 & 1.33 & 3.78 & 5.46\\
        \hline
        35 keV & 500 & 3.25 & 2.62 & 4.99 & 5.94 & 8.45\\
        35 keV & 400 & 2.40 & 0.76 & 3.55 & 4.32 & 6.24\\
        35 keV & 350 & 2.10 & 0.21 & 1.82 & 3.78 & 5.46\\
        \hline
        25 keV & 600 & 4.25 & 2.97 & 7.95 & 7.65 & 11.05\\
        25 keV & 500 & 3.15 & 1.86 & 5.74 & 5.67 & 8.19\\
        \hline
        \label{Table S2}
    \end{longtable}
\end{ThreePartTable}


\begin{figure}[ht!]
  \includegraphics[width=0.85\textwidth]{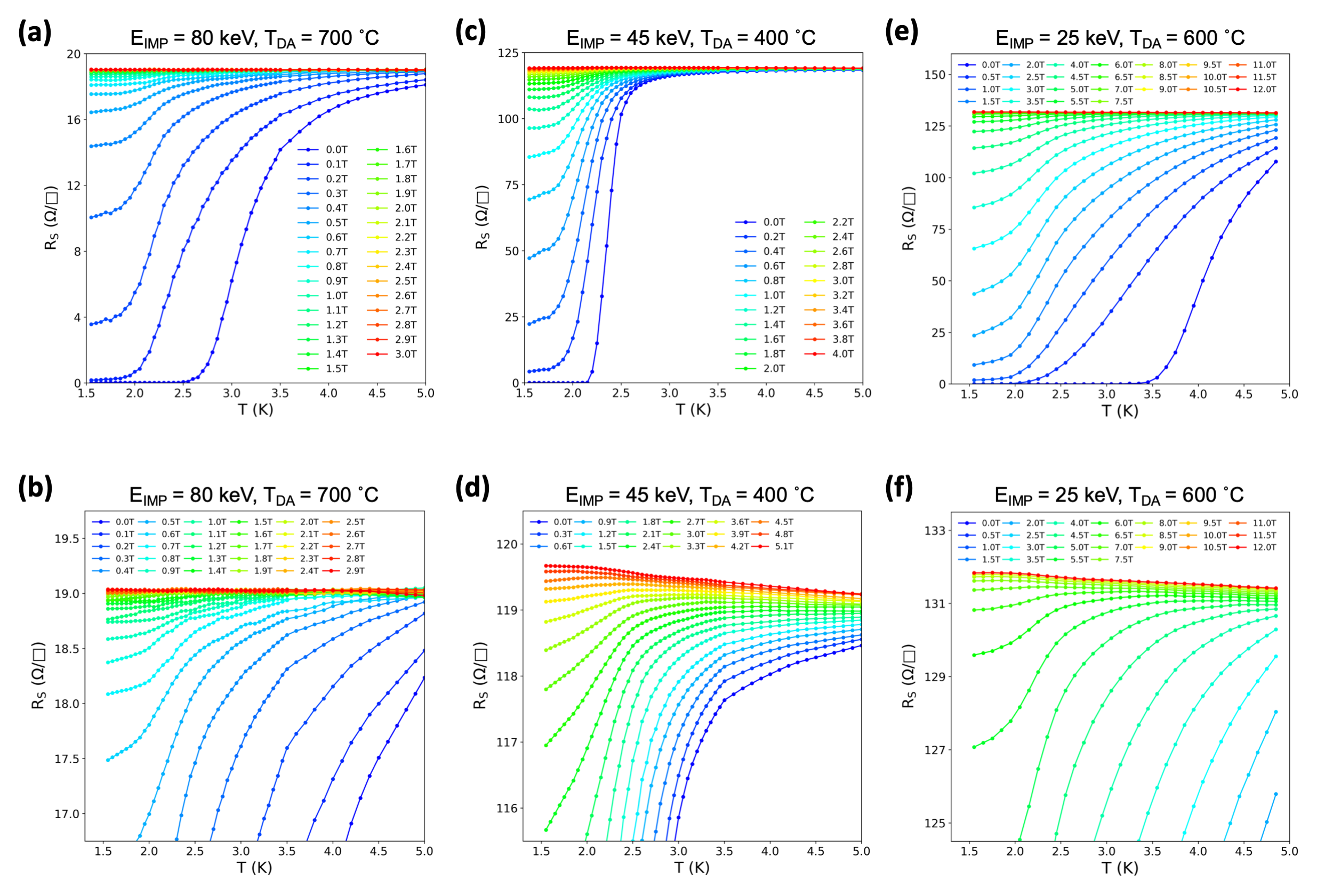}
  \caption{R$_{\rm s}$(T) behavior vs. out-of-plane magnetic field for three samples: (a,b) E$_{\rm IMP}$ = 80keV \& T$_{\rm DA}$ = 700 $^{\circ}$C; (c,d) E$_{\rm IMP}$ = 45keV \& T$_{\rm DA}$ = 400 $^{\circ}$C; (e,f) E$_{\rm IMP}$ = 25keV \& T$_{\rm DA}$ = 600 $^{\circ}$C. In (b), (d), and (e) are the same set of curves are shown over a narrow R$_{\rm s}$ range to emphasize the slope dR/dT vs. B.}
  \label{Fig.S15}
\end{figure}

\clearpage
\section*{Finite-size Scaling}

\begin{figure}[ht!]
  \includegraphics[width=0.75\textwidth]{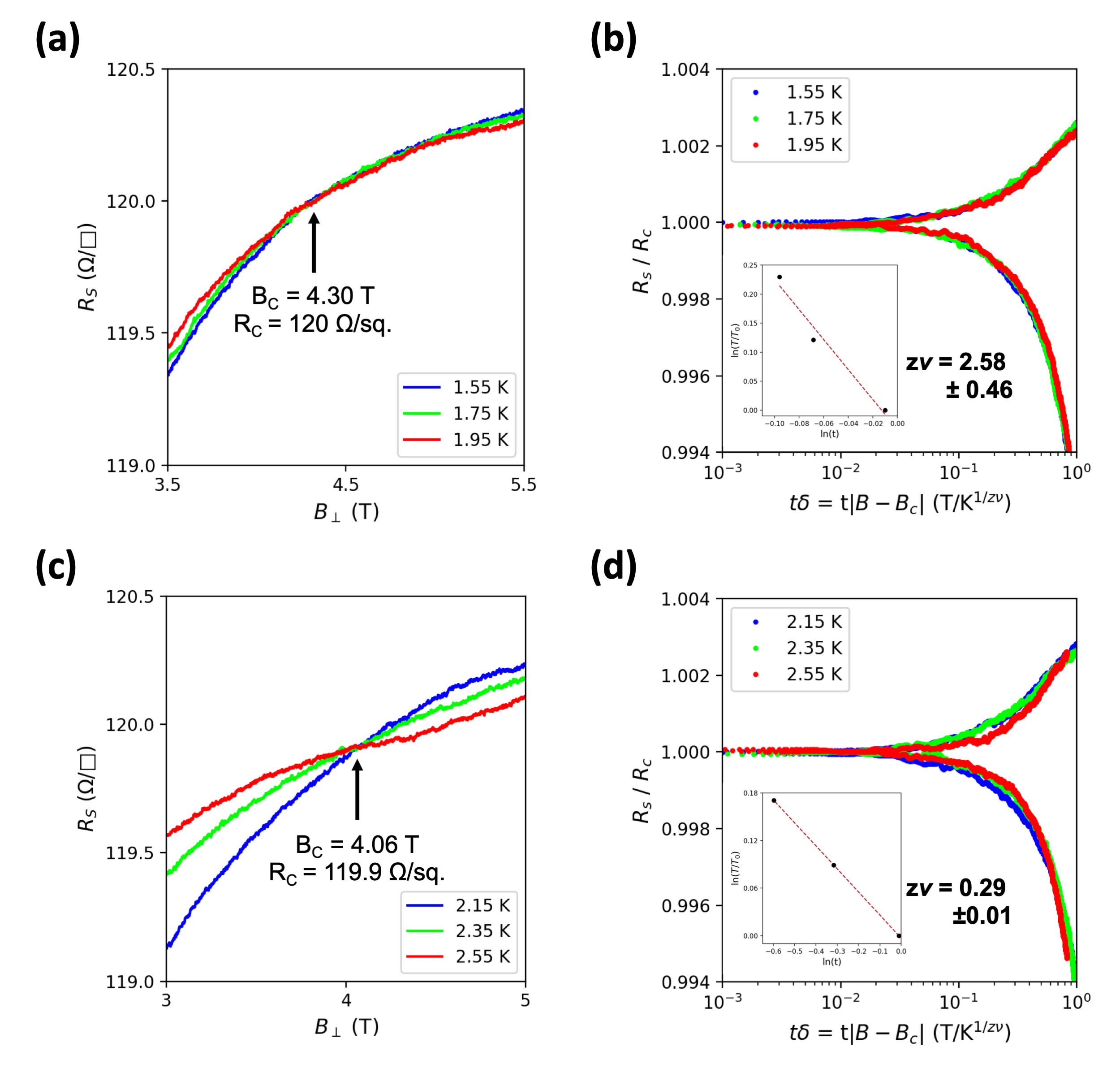}
  \caption{Finite-size scaling analysis for two temperature ranges for the Germanium sample with E$_{\rm IMP}$ = 45 keV \& T$_{\rm DA}$ = 400 $^{\circ}$ C subjected to out-of-plane magnetic field.}
  \label{Fig.S16}
\end{figure}

\begin{figure}[ht!]
  \includegraphics[width=0.7\textwidth]{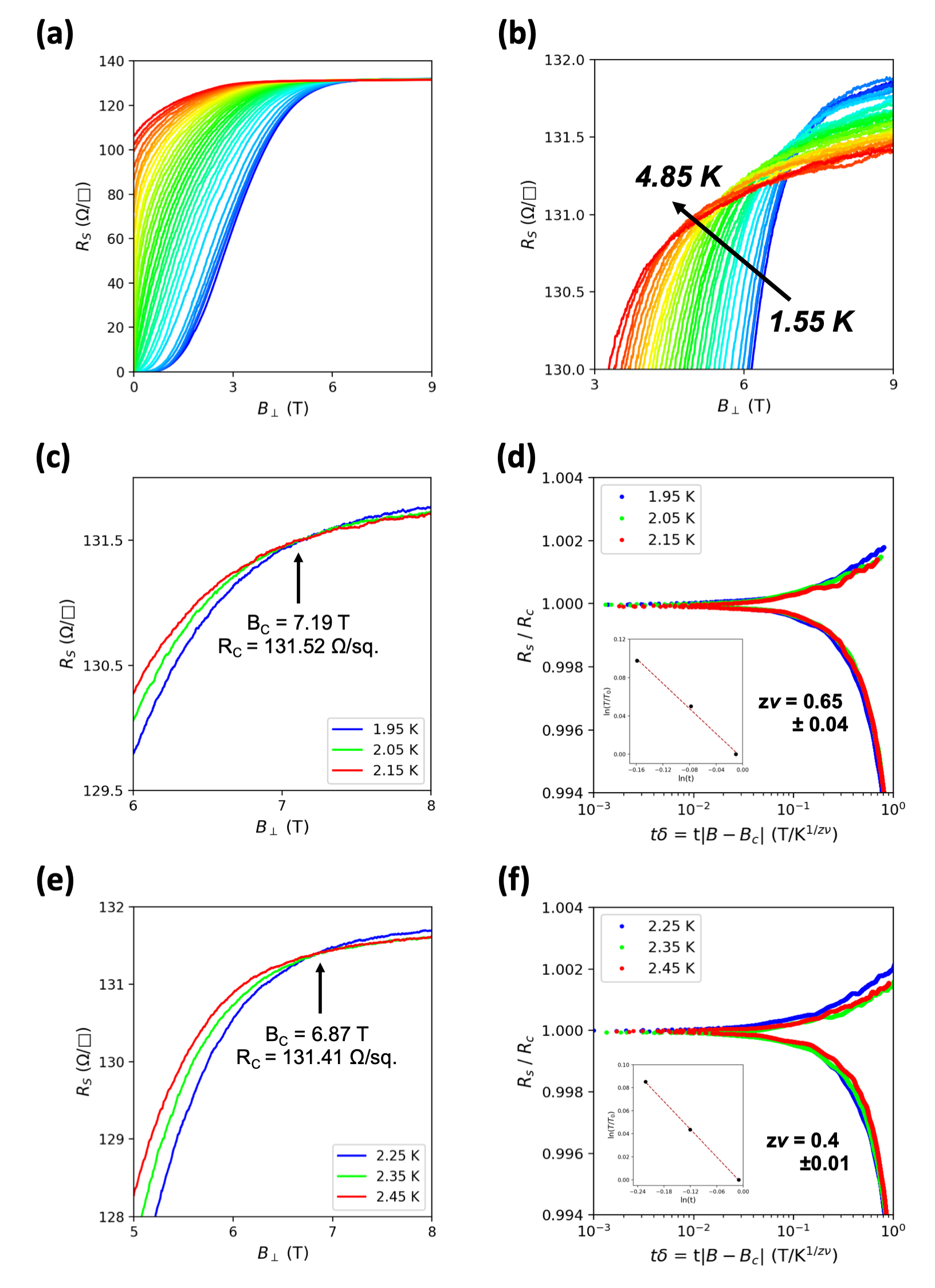}
  \caption{Finite-size scaling analysis for two temperature ranges for the Germanium sample with E$_{\rm IMP}$ = 25 keV \& T$_{\rm DA}$ = 600 $^{\circ}$C subjected to in-plane magnetic field.}
  \label{Fig.S17}
\end{figure}

Finite-size scaling (FSS) analysis was performed to determine the critical exponent product ($z\nu$) for the two samples with prominent magnetoresistance crossing features with E$_{\rm IMP}$ = 45keV \& T$_{\rm DA}$ = 400 $^{\circ}$C (figure \ref{Fig.S16}) and E$_{\rm IMP}$ = 25keV \& T$_{\rm DA}$ = 600 $^{\circ}$C (figure \ref{Fig.S17}). The sheet resistance (R$_s$) for these systems in the zero-temperature quantum critical regime is expected to follow the scaling law of

\begin{equation}
    R_s(\delta, T) = R_c f(\delta T^{\frac{-1}{z\nu}})
    \label{EqScaling}
\end{equation}

where $\delta$ = B - B$_c$ is the offset relative to the critical magnetic field (B$_c$), and R$_c$ is the critical sheet resistance. These values are critical parameters for the quantum phase transition (QPT). Therefore, they should not be confused with superconductor-insulator transition critical parameters discussed earlier in the paper. Moreover, $f$ is an arbitrary scaling function that satisfies $f(0) = 1$, $\nu$ is the coherence length exponent, and $z$ is the dynamical critical exponent \cite{markovic_thickness--magnetic_1998, fisher_boson_1989}. We determined the effective $z\nu$ exponent products for two crossing regions (i.e. 1.55--1.75K, 2.15--2.55K) in the R$_s$(B) isotherms. Each temperature region, which included three crossing curves, was assigned a critical point with characteristic B$_c$ and R$_c$. We then defined rewrote equation \ref{EqScaling} as  R$_s$($\delta, t$) = R$_c$$f(\delta t)$, where $t \equiv (T/T_0)^{-1/z\nu}$ and $T_0$ is the lowest temperature in the critical region (i.e. 1.55K \& 2.15K). The parameter, $t$ is then found by numerically collapsing other R$_s$(B) curves onto the curve of lowest temperature. We finally determine the $z\nu$ exponent the $t$ vs. $T$ in each region to a power law in temperature. This is show in insets for figure \ref{Fig.S16}b \& d and figure \ref{Fig.S17}d \& f.


\end{document}